\documentclass[prd,showpacs,floatfix,amsmath,amssymb,nofootinbib,twocolumn]{revtex4-2}
\usepackage{graphicx,color,dcolumn,booktabs,bm}
\usepackage{longtable,braket}
\usepackage{times}
\usepackage{overpic}
\usepackage{amssymb,tipa}
\usepackage{indentfirst}
\usepackage{feynmf}   
\usepackage{slashed}  
\usepackage{cases}
\usepackage{psfrag}
\usepackage{subfigure}
\usepackage{color}
\usepackage{multirow}
\usepackage{lineno}
\usepackage{changes}

\usepackage[colorlinks, citecolor=blue,anchorcolor=red,menucolor=red,linkcolor=blue,filecolor=red,runcolor=red,urlcolor=blue,frenchlinks=red]{hyperref}
\usepackage{cleveref}

\usepackage{ulem}
\def\be{\begin{equation}}
\def\ee{\end{equation}}

\crefformat{figure}{Fig.~#2#1#3}

\newcommand{\colrui}[1]{\textcolor{cyan}{#1}}

\begin{document}

\title{The rotation effect on the thermodynamics of the QCD matter}
\author{Fei Sun$^{a,b,c}$}\email{ sunfei@ctgu.edu.cn (Corresponding author)}
\author{Shuang Li$^{a,c}$}\email{ 
lish@ctgu.edu.cn}
\author{Rui Wen$^{b}$}\email{ 
rwen@ucas.ac.cn}
\author{Anping Huang$^{b}$}\email{ 
 huanganping@ucas.ac.cn}
\author{Wei Xie$^{a,c}$}\email{ 
xiewei@ctgu.edu.cn}

\affiliation{$^a$Department of Physics, China  Three Gorges University, Yichang, 443002, China \\
$^b$School of Nuclear Science and Technology, University of Chinese Academy of Sciences, Beijing 100049, China\\
$^c$Center for Astronomy and Space Sciences, China Three Gorges University, Yichang 443002, China}

\begin{abstract}
In this study, we investigate the impact of rotation on the thermodynamic characteristics of QCD matter using the three-flavor NJL model. We examine the temperature, quark chemical potential, and angular velocity dependencies of key thermodynamic quantities, such as the trace anomaly, specific heat, speed of sound, angular momentum, and moment of inertia. As the main finding of our analysis, we observe that the speed of sound exhibits a nonmonotonic behavior as the angular velocity changes.
\end{abstract}

\keywords{Rotating quark matter, NJL model, Thermodynamics.}


\maketitle

\section{Introduction\label{sec1}}

Over the decades, intensive investigations in high-energy physics have revealed the rich variety of phenomena exhibited by quantum chromodynamics (QCD) matter at finite temperatures and/or baryon densities. Consequently, determining the phase diagram has become a topic of considerable interest in this field, as it is shaped by the fundamental properties of QCD, namely spontaneous chiral symmetry breaking and confinement. The QCD diagram plays a vital role not only in understanding heavy-ion collision experiments but also in shedding light on the early universe and compact stars. The critical endpoint (CEP) on the phase diagram offers crucial information for inferring the phase boundary, which can be characterized by critical phenomena that manifest in thermodynamic and hydrodynamic properties. Therefore, exploring the thermodynamics of strong interaction matter contributes significantly to understanding the phase diagram and the properties of matter created in heavy-ion experiments. Extensive research has already examined the phase transition in the presence of finite temperature and chemical potential.

In recent years, there has been a shift in focus towards noncentral high-energy heavy-ion collisions (HIC), where both a strong magnetic field and rotation are generated. The study of matter under these extreme conditions is of great interest in the field of QCD. The QCD matter created through off-central collisions carries a nonzero angular momentum on the order of $10^{4} \sim 10^{5}\hbar$ with local angular velocity ranging from $0.01\sim 0.1$ GeV \cite{Liang:2004ph,Huang:2011ru,Becattini:2007sr,Csernai:2013bqa,Deng:2016gyh,Jiang:2015cva}, and the angular momentum affects both the orbital motion and the individual spins of the particles. Among them, it is particularly important to investigate the phase diagram and thermodynamics of QCD matter under rotation through experimental, theoretical, and computational studies. In 2017, the STAR Collaboration published the first observation of global polarization resulting from noncentral heavy-ion collisions, which has led to the exploration of numerous spin-related quantum phenomena and the remarkably strong fluid vorticity structures\cite{STAR:2017ckg,STAR:2018gyt,STAR:2019erd}. Additionally, several upcoming experiments, the Facility for Antiproton and Ion Research (FAIR) in Darmstadt, the Nuclotron-based Ion Collider Facility (NICA) in Dubna, and the CSR-External target Experiment (CEE) in Huizhou are also planning to conduct BES to identify the CEP, and also to conduct noncentral heavy-ion collision experiments.

In theoretical aspects, there have also been significant advances on the matter under rotation.
The influence of rotation on various physical scenarios is currently being actively investigated. These include noncentral heavy-ion collisions in high-energy nuclear physics \cite{Becattini:2015ska,Jiang:2016woz,Shi:2017wpk,Pang:2016igs,Becattini:2016gvu,Xia:2018tes}, as well as in hadron physics \cite{Zhang:2018ome,Xu:2022hql}, trapped non-relativistic bosonic cold atoms in condensed matter physics \cite{Fetter:2009zz,PhysRevA.78.011601,PhysRevA.79.053621,Gooth:2017mbd}, and rapidly spinning neutron stars in astrophysics \cite{Watts:2016uzu,Grenier:2015pya,10.1111/j.1365-2966.2005.08812.x}. Due to the non-Abelian nature of QCD, a thorough understanding of dynamical chiral symmetry breaking (DCSB) is challenging, several effective models have been proposed, such as the Nambu and Jona-Lasinio (NJL) models, quark-meson (QM) models, holographic QCD models; also some functional QCD approaches have been developed, such as Dyson-Schwinger equations (DSE), functional renormalization group (fRG), as well as their extensions to investigate the QCD phase diagram \cite{Asakawa:1989bq,Stephanov:1998dy,Stephanov:1999zu,Fukushima:2003fw,Megias:2004hj,Ratti:2005jh,Schaefer:2007pw,Skokov:2010sf,Skokov:2010wb,Herbst:2010rf,Drews:2016wpi,Fukushima:2017csk,Klevansky:1992qe,Zhuang:1994dw,Buballa:2003qv,Kohyama:2015hix,Roberts:1994dr,Alkofer:2000wg,Fischer:2006ub,Cloet:2013jya,Schwarz:1999dj,Zhuang:2000ub,Chen:2014ufa,Chen:2015dra,Fan:2016ovc,Fan:2017mrk,Fu:2010ay,Bowman:2008kc,Mao:2009aq,Schaefer:2011ex,Schaefer:2012gy,Qin:2010nq,Luecker:2013oda,Fu:2016tey,Chen:2017lsf,Fu:2019hdw,Wen:2018nkn,Mamani:2020pks,Chen:2020ath,Sun:2023dwh,Lu:2023msn}. During a phase transition, thermodynamic quantities may undergo abrupt changes or exhibit highly unusual behavior. Therefore, describing the QCD transition requires determining thermodynamic quantities such as pressure, energy density, entropy density, trace anomaly, specific heat, speed of sound, and when considering rotating systems, additional quantities such as angular momentum and moment of inertia need to be determined. Therefore, it is natural to investigate how rotation affects the phase transition and thermodynamics.

Computationally, the QCD phase diagram has been investigated using first-principal Lattice QCD (LQCD) calculations, which have been successful in studying the system at zero chemical potential. However, when dealing with large chemical potentials, the predictions made by LQCD, even with the use of Taylor expansion of the partition function, still have certain limitations \cite{deForcrand:2009zkb}. In rotating systems, the angular velocity plays a similar role to an effective chemical potential, therefore, in LQCD simulations of rotating systems, the sign problem is also encountered. Nevertheless, there have been some advancements in Lattice simulations of rotating systems  \cite{Braguta:2022str,Chernodub:2022veq,Braguta:2023yjn,Braguta:2023kwl,Yang:2023vsw}.

It has been highlighted in Ref. \cite{Fujimoto:2020tjc} that accounting for the contributions from finite strange quark mass is crucial when studying realistic environments of neutron stars. In particular, compared to light quarks, determining the phase diagram of strange quark matter has become a subject of significant theoretical and experimental endeavors. This is due to the crucial role played by the strange quark in shaping the behavior of the phase diagram for both chiral and deconfinement transitions. Furthermore, the strange quark also has a profound impact on the stability limit of neutron stars, which are believed to exist under extreme temperature and pressure conditions \cite{Avancini:2012ee}. Interest in investigating neutron stars has surged following the remarkable findings of the LIGO and VIRGO collaborations \cite{PhysRevLett.119.161101}. Additionally, since compact stars like neutron stars can exhibit rapid rotation, exploring the effects of rotation on the phase transitions of these astrophysical objects is both intriguing and significant. So, the extension to $2+1$ flavors with inclusion of strange quarks should be explored and there are several works on this topic \cite{Fu:2007xc,Borsanyi:2013bia,Sun:2021hxo}.

Investigating QCD matter under rotation is a captivating area of research. In addition to studying transport properties like the chiral vortical effect and chiral vortical wave  \cite{Kharzeev:2007tn,Son:2009tf,Kharzeev:2010gr,Kharzeev:2015znc}, it is also highly intriguing to explore the effects of rotation on phase transitions and thermodynamics. The QCD phase diagram is expected to exhibit great complexity in the presence of angular velocity and chemical potential, potentially revealing interesting phases and regimes. Studying the thermodynamics of quarks (including the strange quark), under extreme conditions such as large chemical potential and strong rotation, can contribute to our understanding of compact stars, the evolution of the universe, and high-energy nuclear physics. Thermodynamics has been the subject of intense investigation in recent years, it is natural to inquire about the influence of rotation on QCD thermodynamics. In this paper, our main focus is the global rotation effect on the thermodynamics of the three-flavor NJL model. We will also explore the model's capacity to describe the essential aspects of QCD thermodynamics around the critical point.

Our work is organized as follows. In \Cref{sec2}, we begin by introducing the formalism of the three-flavor NJL model and derive the detailed expressions for thermodynamics in the presence of rotation. In \Cref{sec3}, we present the numerical results and discussions on thermodynamic quantities, analyzing both the light quarks and strange quark. Finally, in \Cref{sec4}, we summarize our findings and conclude the paper.

\section{FORMALISM\label{sec2}}
First, we provide a very brief sketch of the basis for studying the rotating matter. The metric tensor can describe the structure of space-time under rotating frame reads\\
\begin{eqnarray}
g_{\mu\nu}=\left(
\begin{array}{cccc}
 1-{\vec v}^{\, 2} & -v_1 & -v_2 & -v_3 \\
 -v_1 & -1 & 0 & 0 \\
 -v_2 & 0 & -1 & 0 \\
 -v_3 & 0 & 0 & -1 \\
\end{array}
\right),
\end{eqnarray}\\
where $v_i$ is the velocity.
Our starting point is the partition function
\begin{eqnarray}
{\cal Z} = \int {D[\bar \psi ]} D[\psi ]e^{iS},
\end{eqnarray}
here, $S$ denotes the  quark action, which is the integration of the Lagrangian density $\mathcal{L}$. When extending to the case of rotating fermions  \cite{Fetter:2009zz,Jiang:2016wvv,Yamamoto:2013zwa} with non-zero chemical potential, the Lagrangian in the three-flavor NJL model is given by

\begin{align}
 {\cal L} & = \bar\psi \left( {i\bar{\gamma} ^\mu (\partial _\mu+\Gamma_{\mu})- m+\gamma ^0 \mu } \right)\psi \nonumber\\
   & + G\sum\limits_{a = 0}^8 {\left( {\bar \psi \lambda ^a \psi } \right)^2 } \nonumber \\
   & -K\{ \rm{det}[\bar \psi (1+\gamma^{5})\psi] +\rm{det}[\bar \psi (1-\gamma^{5})\psi]  \},
\end{align}
here, $\psi$ is the quark field, $\bar{\gamma}^\mu=e_{a}^{\ \mu} \gamma^a$ with $e_{a}^{\ \mu}$ being the tetrads for spinors and $\gamma^{a}$ represents the gamma matrix, $\Gamma_\mu$ is defined as $\Gamma_\mu=\frac{1}{4}\times\frac{1}{2}[\gamma^a,\gamma^b] \ \Gamma_{ab\mu}$\colrui{,} which is the spinor connection, where $\Gamma_{ab\mu}=\eta_{ac}(e^c_{\ \sigma} G^\sigma_{\ \mu\nu}e_b^{\ \nu}-e_b^{\ \nu}\partial_\mu e^c_{\ \nu})$, and $G^\sigma_{\ \mu\nu}$ is the affine connection determined by $g^{\mu\nu}$ \colrui{.} $m$ is the bare quark mass matrix, $\mu$ denotes the chemical potential, and $G$ represents the coupling constant of four-point interaction term. $\lambda ^a(a=1,...8) $ are the Gell-Mann matrices in flavor space. The last term corresponds to the t'Hooft interaction with coupling strength $K$,
which is a determinant in flavor space. Considering a system with an angular velocity along the fixed $z$-axis, then $\vec{v}=\vec{\omega}\times \vec{x}$. By choosing $e^{a}_{\ \mu}=\delta^a_{\ \mu}+  \delta^a_{\ i}\delta^0_{\ \mu} \, v_i$ and $e_{a}^{\ \mu}=\delta_a^{\ \mu} -  \delta_a^{\ 0}\delta_i^{\ \mu} \, v_i$, and expanding to first order of angular velocity the Lagrangian has following expression:
\begin{eqnarray}
{\cal L} &=& \bar \psi \left[ {i{\gamma ^\mu }{\partial _\mu } - m + {\gamma ^0}\mu } \right]\psi \nonumber \\ \nonumber
\\
 &+& \bar \psi \left[ {{{\left( {{\gamma ^0}} \right)}^{ - 1}}\left( {\left( {\mathord{\buildrel{\lower3pt\hbox{$\scriptscriptstyle\rightharpoonup$}} 
\over \omega }  \times \mathord{\buildrel{\lower3pt\hbox{$\scriptscriptstyle\rightharpoonup$}} 
\over x} } \right) \cdot \left( { - i\mathord{\buildrel{\lower3pt\hbox{$\scriptscriptstyle\rightharpoonup$}} 
\over \partial } } \right) + \mathord{\buildrel{\lower3pt\hbox{$\scriptscriptstyle\rightharpoonup$}} 
\over \omega } \cdot  {{\mathord{\buildrel{\lower3pt\hbox{$\scriptscriptstyle\rightharpoonup$}} 
\over S} }_{4 \times 4}}} \right)} \right]\psi  \nonumber\\ \nonumber
\\
 &+& G\sum\limits_{a = 0}^8 {{{\left( {\bar \psi {\lambda ^a}\psi } \right)}^2}} \nonumber \\ \nonumber
\\
 &-& K\{ {\rm{det}}[\bar \psi (1 + {\gamma ^5})\psi ] + {\rm{det}}[\bar \psi (1 - {\gamma ^5})\psi ], \label{Diracequation}
\end{eqnarray}
where $\mathord{\buildrel{\lower3pt\hbox{$\scriptscriptstyle\rightharpoonup$}}
\over S} _{4 \times 4}  = \frac{1}{2}\left( {\begin{array}{*{20}c}
   {\mathord{\buildrel{\lower3pt\hbox{$\scriptscriptstyle\rightharpoonup$}}
\over \sigma } } & 0  \\
   0 & {\mathord{\buildrel{\lower3pt\hbox{$\scriptscriptstyle\rightharpoonup$}}
\over \sigma } }  \\
\end{array}} \right)$ is the spin operator. With the technique of the path integral formulation for Grassmann variables theory and the mean field approximation, the linearization is done to a 4-quark interaction and 6-quark interaction, we get the expression of $\log {\cal Z}$ as follows:

\begin{eqnarray}
 \log {\cal Z}&=& \frac{1}{T}\int {{d^3}x} \left( {2G\sum\limits_f {{{\left\langle {{{\bar \psi }_f}{\psi _f}} \right\rangle }^2}}  - 4K\prod\limits_f {\left\langle {{{\bar \psi }_f}{\psi _f}} \right\rangle } } \right) \nonumber \\
 &+& \sum\limits_f {\log } \det \frac{{D_f^{ - 1}}}{T}.\label{logz}
\end{eqnarray}
The inverse fermion propagator ${D^{ - 1} }$ in Eq. (\ref{logz}) can be derived as follows,
\begin{eqnarray}
D^{ - 1}  = \gamma ^0 \left( { - i{\boldsymbol{{\omega_{l}}}}  + \left( {n + \frac{1}{2}} \right)\omega  + \mu } \right) - M - \mathord{\buildrel{\lower3pt\hbox{$\scriptscriptstyle\rightharpoonup$}}
\over \gamma } .\mathord{\buildrel{\lower3pt\hbox{$\scriptscriptstyle\rightharpoonup$}}
\over p},
\end{eqnarray}
here we introduce Matsubara frequency ${\boldsymbol{{\omega_{l}}}}=-ip_{0}=2{\pi}lT$ with the temperature $T$, and $M$ denotes the dynamical mass of quark
\begin{eqnarray}
{M_q} &=& {m_q} + \left( {2K\left\langle {\bar ss} \right\rangle  - 4G} \right)\left\langle {\bar qq} \right\rangle,
\\ \nonumber
\\
{M_s} &=& {m_s} - 4G\left\langle {\bar ss} \right\rangle  + 2K{\left\langle {\bar qq} \right\rangle ^2}\;.
\end{eqnarray}

To find solutions of the Dirac equation, we start by choosing a complete set of commutating operators consisting of $\hat{H}$, which can be obtained from Eq. (\ref{Diracequation}) by using the relation ${\cal H} = \bar \psi \left( {i\gamma ^0  \partial _0} \right)\psi-{\cal L}$, the momentum in the $z$-direction $\hat{p}_{z}$, the square of transverse momentum $\hat{p}_{t}^{2}$, the $z$-component of the total angular momentum $\hat{J}_{z}$ and the transverse helicity $\hat{h}_{t}$, here ${\hat h_t} = {\gamma ^5}{\gamma ^3}{\mathord{\buildrel{\lower3pt\hbox{$\scriptscriptstyle\rightharpoonup$}}
\over p} _t} \cdot \mathord{\buildrel{\lower3pt\hbox{$\scriptscriptstyle\rightharpoonup$}}
\over S} $, and ${\gamma ^5}=i{\gamma ^0}{\gamma ^1}{\gamma ^2}{\gamma ^3}$. By solving the eigenvalue equations of the complete set of commuting operators \{$\hat{H}$, $\hat{p}_{z}$, $\hat{p}_{t}^{2}$,$\hat{J}_{z}$, $\hat{h}_{t}$\}, we obtain the positive and negative energy solutions of the Dirac field as follows:
In cylindrical coordinates, the general spinor eigenstates can be written as

\begin{eqnarray}
u = \sqrt {\frac{{E + m}}{4E}} \left( {\begin{array}{*{20}c}
   {e^{ip_z z} e^{i{\rm{n}}\theta } J_n \left( {p_t r} \right)}  \\
   {se^{ip_z z} e^{i\left( {n + 1} \right)\theta } J_{n + 1} \left( {p_t r} \right)}  \\
   {\frac{{p_z  - isp_t }}{{E + m}}e^{ip_z z} e^{i{\rm{n}}\theta } J_n \left( {p_t r} \right)}  \\
   {\frac{{ - sp_z  + ip_t }}{{E + m}}e^{ip_z z} e^{i\left( {n + 1} \right)\theta } J_{n + 1} \left( {p_t r} \right)}  \\
\end{array}} \right),
\end{eqnarray}
\begin{eqnarray}
v = \sqrt {\frac{{E + m}}{4E}} \left( {\begin{array}{*{20}c}
   {\frac{{p_z  - isp_t }}{{E + m}}e^{ - ip_z z} e^{i{\rm{n}}\theta } J_n \left( {p_t r} \right)}  \\
   {\frac{{ - sp_z  + ip_t }}{{E + m}}e^{ - ip_z z} e^{i\left( {n + 1} \right)\theta } J_{n + 1} \left( {p_t r} \right)}  \\
   {e^{ - ip_z z} e^{i{\rm{n}}\theta } J_n \left( {p_t r} \right)}  \\
   { - se^{ - ip_z z} e^{i\left( {n + 1} \right)\theta } J_{n + 1} \left( {p_t r} \right)}  \\
\end{array}} \right).
\end{eqnarray}
Here, $s=\pm1$ represent the transverse helicity, $n$ denotes the $z$-direction angular momentum quantum number. After the summation of all the Matsubara frequencies and carrying out the general approach of the finite temperature fields \cite{kapusta_gale_2006}, it can be shown that the grand thermodynamic potential ($\Omega=-\frac{T}{V} \log {\cal Z}$) has following form,
\begin{widetext}
\begin{eqnarray}
 \begin{array}{l}
\Omega {\rm{ = }}2G\left( {2{{\left\langle {\bar qq} \right\rangle }^2} + {{\left\langle {\bar ss} \right\rangle }^2}} \right) - 4K{\left\langle {\bar qq} \right\rangle ^2}\left\langle {\bar ss} \right\rangle \\
\\
 - \frac{3}{{2{\pi ^2}}}\sum\limits_{n =  - \infty }^\infty  {\int_0^\Lambda  {{{\rm{p}}_t}d{{\rm{p}}_t}\int_{ - \sqrt {{\Lambda ^2} - p_t^2} }^{\sqrt {{\Lambda ^2} - p_t^2} } {d{p_z}} } \left( {\left( {{J_{n + 1}}{{({{\rm{p}}_t}r)}^2} + {J_n}{{({{\rm{p}}_t}r)}^2}} \right)} \right.} \left( {\varepsilon _{q}  - \left( {\frac{1}{2} + n} \right)\omega } \right)\\
\\
 - \frac{3}{{2{\pi ^2}}}\sum\limits_{n =  - \infty }^\infty  {\int_0^{\infty} {{{\rm{p}}_t}d{{\rm{p}}_t}\int_{ -\infty }^{\infty } {d{p_z}} } \left( {\left( {{J_{n + 1}}{{({{\rm{p}}_t}r)}^2} + {J_n}{{({{\rm{p}}_t}r)}^2}} \right)} \right.} T\left\{ {\log \left( {{e^{ - \frac{{ - {\mu _q} + \varepsilon _{q}  - \left( {\frac{1}{2} + n} \right)\omega }}{T}}} + 1} \right)} \right.\\
\\
\left. { + \log \left( {{e^{ - \frac{{{\mu _q} + \varepsilon _{q}  - \left( {\frac{1}{2} + n} \right)\omega }}{T}}} + 1} \right)} \right\}\\
\\
 - \frac{3}{{4{\pi ^2}}}\sum\limits_{n =  - \infty }^\infty  {\int_0^{{\Lambda}} {{{\rm{p}}_t}d{{\rm{p}}_t}\int_{ - \sqrt {{\Lambda ^2} - p_t^2} }^{\sqrt {{\Lambda ^2} - p_t^2} } {d{p_z}} } \left( {\left( {{J_{n + 1}}{{({{\rm{p}}_t}r)}^2} + {J_n}{{({{\rm{p}}_t}r)}^2}} \right)} \right.} \left( {\varepsilon _{s}  - \left( {\frac{1}{2} + n} \right)\omega } \right)\\
\\
 - \frac{3}{{4{\pi ^2}}}\sum\limits_{n =  - \infty }^\infty  {\int_0^{\infty} {{{\rm{p}}_t}d{{\rm{p}}_t}\int_{ - \infty }^{\infty } {d{p_z}} } \left( {\left( {{J_{n + 1}}{{({{\rm{p}}_t}r)}^2} + {J_n}{{({{\rm{p}}_t}r)}^2}} \right)} \right.} T\left\{ {\log \left( {{e^{ - \frac{{ - {\mu _s} + \varepsilon _{s}  - \left( {\frac{1}{2} + n} \right)\omega }}{T}}} + 1} \right)} \right.\\
\\
\left. { + \log \left( {{e^{ - \frac{{{\mu _s} + \varepsilon _{s}  - \left( {\frac{1}{2} + n} \right)\omega }}{T}}} + 1} \right)} \right\}.
\end{array}
\end{eqnarray}
\end{widetext}
Here, the quark quasiparticle energy
${\varepsilon } = \sqrt {M^2 + p_t^2 + p_z^2}$. For simplicity we also introduce the quark quasiparticle energy under rotation as follows,
\begin{align}
{\varepsilon _{f, n}} = {\varepsilon _{f}}- \left( {\frac{1}{2} + n} \right)\omega.
\end{align}
Note that, the above expression of grand thermodynamic potential contains an explicit cutoff dependence, due to the NJL model being nonrenormalizable. Here, our thermodynamic potential naturally separates into the vacuum piece and the temperature-dependent matter part, which is very helpful in calculating the thermodynamic quantities. The three-momentum cutoff in the vacuum piece should be chosen to reproduce observables, such as pion mass, pion decay constant, and so on, also, in principle, the cutoff in the matter part should take the infinite value. Here, $\Lambda$ is the three-momentum cutoff of the vacuum part in the potential.

Then, we consider the gap equations which will be required to minimize the grand potential, the dynamical quark mass $M_{f}$ can be determined by solving the stationary condition, and we also require the solutions satisfy to get the minimum of the potential, namely, 
\begin{eqnarray}
\frac{{\partial \Omega }}{{\partial \left\langle {\bar qq} \right\rangle }} = \frac{{\partial \Omega }}{{\partial \left\langle {\bar ss} \right\rangle }} = 0,
\end{eqnarray}
\begin{eqnarray}
\frac{{\partial ^2 \Omega }}{{{\partial \left\langle {\bar qq} \right\rangle }^2 }}>0, ~~\frac{{\partial ^2 \Omega }}{{\partial \left\langle {\bar ss} \right\rangle^2 }}>0,
\end{eqnarray}
which leads to the following coupled gap equations:
\begin{widetext}
\begin{eqnarray}
0 &=& \left( {8G\left\langle {\bar qq} \right\rangle  - 8K\left\langle {\bar ss} \right\rangle \left\langle {\bar qq} \right\rangle } \right) - \frac{3}{{{\pi ^2}}}\sum\limits_{n =  - \infty }^\infty  {\int_0^\Lambda  {{{\rm{p}}_t}d{{\rm{p}}_t}\int_{ - \sqrt {{\Lambda ^2} - p_t^2} }^{\sqrt {{\Lambda ^2} - p_t^2} } {d{p_z}} } \left( {\left( {{J_{n + 1}}{{({{\rm{p}}_t}r)}^2} + {J_n}{{({{\rm{p}}_t}r)}^2}} \right)} \right.} \nonumber\\ \nonumber 
\\ 
 &\times& \left( {\frac{{\left( { - 2G + K\left\langle {\bar ss} \right\rangle } \right){M_q}}}{\varepsilon_{u}} + \frac{{K\left\langle {\bar qq} \right\rangle M_s}}{{\varepsilon_{s}}}} \right)\nonumber\\ \nonumber  
\\ 
 &+& \frac{3}{{{\pi ^2}}}\sum\limits_{n =  - \infty }^\infty  {\int_0^{\infty} {{{\rm{p}}_t}d{{\rm{p}}_t}\int_{ - \infty }^{\infty} {d{p_z}} } \left( {\left( {{J_{n + 1}}{{({{\rm{p}}_t}r)}^2} + {J_n}{{({{\rm{p}}_t}r)}^2}} \right)} \right.} \nonumber\\ \nonumber
\\ 
 &\times& \left\{ {\frac{\left( { - 2G + K\left\langle {\bar ss} \right\rangle } \right)M_q}{\varepsilon_{q}}\left[n_f(\varepsilon_{q,n},T,\mu)+ \bar n_f(\varepsilon_{q,n},T,\mu)\right]
+ \frac{K\left\langle {\bar qq} \right\rangle M_s}{\varepsilon_{s}}\left[n_f(\varepsilon_{s,n},T,\mu)+ \bar n_f(\varepsilon_{s,n},T,\mu)\right ] } \right\},
\end{eqnarray}
\begin{eqnarray}
0 &=& \left( {4G\left\langle {\bar ss} \right\rangle  - 4K{{\left( {\left\langle {\bar qq} \right\rangle } \right)}^2}} \right) - \frac{3}{{{\pi ^2}}}\sum\limits_{n =  - \infty }^\infty  {\int_0^\Lambda  {{{\rm{p}}_t}d{{\rm{p}}_t}\int_{ - \sqrt {{\Lambda ^2} - p_t^2} }^{\sqrt {{\Lambda ^2} - p_t^2} } {d{p_z}} } \left( {\left( {{J_{n + 1}}{{({{\rm{p}}_t}r)}^2} + {J_n}{{({{\rm{p}}_t}r)}^2}} \right)} \right.}\nonumber \\ \nonumber
\\
 &\times& \left( {\frac{{K\left\langle {\bar qq} \right\rangle M_q}}{{\varepsilon_{q}}} - \frac{{G{M_s}}}{{\varepsilon_{s}}}} \right)\nonumber\\ \nonumber
\\
 &+& \frac{3}{{{\pi ^2}}}\sum\limits_{n =  - \infty }^\infty  {\int_0^{\infty} {{{\rm{p}}_t}d{{\rm{p}}_t}\int_{ - \infty}^{\infty} {d{p_z}} } \left( {\left( {{J_{n + 1}}{{({{\rm{p}}_t}r)}^2} + {J_n}{{({{\rm{p}}_t}r)}^2}} \right)} \right.} \nonumber\\ \nonumber
\\
 &\times& \left\{ \frac{K\left\langle {\bar qq} \right\rangle M_q}{\varepsilon_{q}}\left[n_f(\varepsilon_{q,n},T,\mu)+ \bar n_f(\varepsilon_{q,n},T,\mu)\right]
-\frac{G M_s}{\varepsilon_{s}}\left[n_f(\varepsilon_{s,n},T,\mu)+ \bar n_f(\varepsilon_{s,n},T,\mu)\right] \right\},
 \end{eqnarray}
\end{widetext}
here, $n_f$ and $\bar n_f$ denote the quark and anti-quark distribution functions:
\begin{align}
n_f(\varepsilon,T,\mu)&=\frac{1}{e^{\frac{\varepsilon-\mu}{T}}+1},\\
\bar n_f(\varepsilon,T,-\mu)&=\frac{1}{e^{\frac{\varepsilon+\mu}{T}}+1}.
\end{align}

This set of coupled equations is then solved for the fields as functions of temperature $T$,  quark chemical potential $\mu$, and angular velocity $\omega$. Now we turn to the thermodynamics of the rotating system,
when we extend to the rotating system, the vorticity should also be considered a further intensive thermodynamic quantity which is necessary for the description of the local fluid, so, some corrections may need to be carried out for the energy density as follows \cite{Landau_1980,Chernodub:2016kxh,Zhao:2022uxc}:
\begin{eqnarray}
\varepsilon  =  - p + Ts + \mu n + \omega J.
\end{eqnarray}
Here, $n$ denotes quark number density, $J$ presents the (polarization) angular momentum density.

From the standard thermodynamic relations, the pressure, and (polarization) angular momentum density (the angular velocity can be regarded as an ``effective chemical potential", similarly, we can define the angular momentum by a derivative with respect to the angular velocity of the grand canonical potential) and the quark number density are given as follows,
\begin{eqnarray}
p = \Omega\left( {T=0, \mu=0, \omega=0 } \right) - \Omega\left( {T, \mu, \omega } \right),
\end{eqnarray}
\begin{eqnarray}
s=  - {\left( {\frac{{\partial \Omega }}{{\partial T}}} \right)_{\mu ,\omega }},
\end{eqnarray}
\begin{eqnarray}
n =  - {\left( {\frac{{\partial \Omega }}{{\partial \mu}}} \right)_{T ,\omega }} ,
\end{eqnarray}
\begin{eqnarray}
J=  - {\left( {\frac{{\partial \Omega }}{{\partial \omega}}} \right)_{T ,\mu }},
\end{eqnarray}\\
note here, to get a physical pressure, we have renormalized the thermodynamical potential, and the subscript represents keeping the chemical potential and angular velocity fixed during taking the partial differentiation, and the trace anomaly can be defined as
\begin{eqnarray}
\Theta=\varepsilon  - 3p.
\end{eqnarray}
For each flavor, the explicit formulae of entropy density and quark number density, and the angular momentum along the $z$-axis are listed in \Cref{app:thermodynamic}.

Once get the expression of the angular momentum, then we can directly obtain the moment of inertia of the rotating system
\begin{eqnarray}
I=\frac{1}{\omega}(-\frac{d\Omega}{d \omega})=\frac{J}{\omega}.
\end{eqnarray}
For the description of the expansion of dense matter created in heavy ion collisions, a fundamental quantity that determines the expansion of hot dense matter is the speed of sound
\begin{eqnarray}
c_{s}^{2}=\frac{dp}{d\epsilon},
\end{eqnarray}
another quantity of interest is the specific heat
\begin{eqnarray}
C_{V}=\frac{d\epsilon}{dT}.
\end{eqnarray}
Here, we don't list the detailed expressions of them, and one can easily get that from the above expressions.\\

\section{Numerical results and discussions \label{sec3}}

\subsection{Dynamical quark mass and chiral transition}
In this section, we will present our numerical results for the dynamical quark mass and chiral transition in the three-flavor Nambu and Jona-Lasinio (NJL) model under rotation. 
In our calculations, the  input parameters in the NJL are the coupling constants $G$, the light quark mass $m_{q}$ (throughout, we ignore isospin breaking effects and work with $m_{u}=m_{d}=m_{q}$), the strange quark mass $m_{s}$ and the three-momentum cutoff $\Lambda$ and the 't Hooft term coupling constant. We use the model parameters reported in Ref. \cite{Kohyama:2016fif}, which have been estimated by the fitting in light of the following observations: $m_{\pi}=138$ MeV, $f_{\pi}=92$ MeV, $m_{K}=495$ MeV and $m_{\eta'}=958$ MeV, the input parameters are as follows: $m_q = 0.005~{\rm{GeV}},~ m_s  = 0.1283~{\rm{GeV}},
 ~G = 3.672~{\kern 1pt} {\rm{GeV}}^{ - 2},~K = 59.628~{\kern 1pt} {\rm{GeV}}^{ - 5},
 ~\Lambda  = 0.6816~{\rm{GeV}}$. Throughout the text, unless otherwise specified, the radius $r$ is taken as $0.1~ {\rm{GeV}}^{ - 1}$.
\begin{figure}
\subfigure[]
{\includegraphics[width=0.45\textwidth]{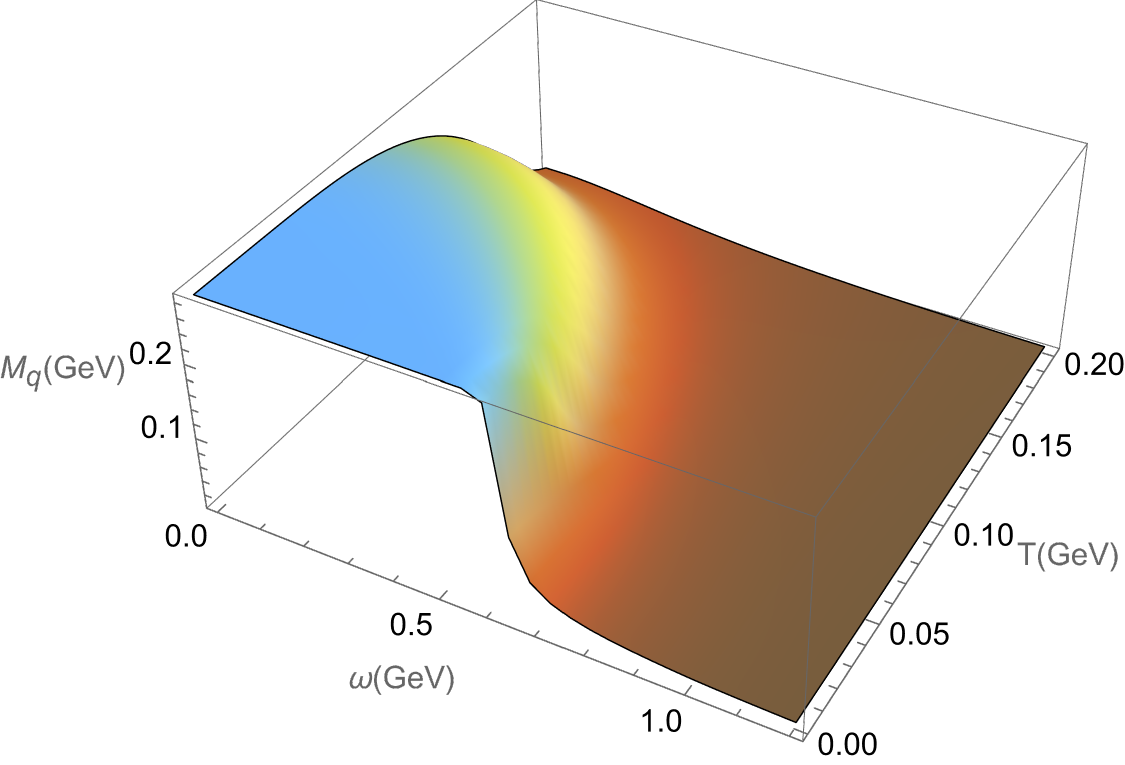}}
\subfigure[]
{\includegraphics[width=0.45\textwidth]{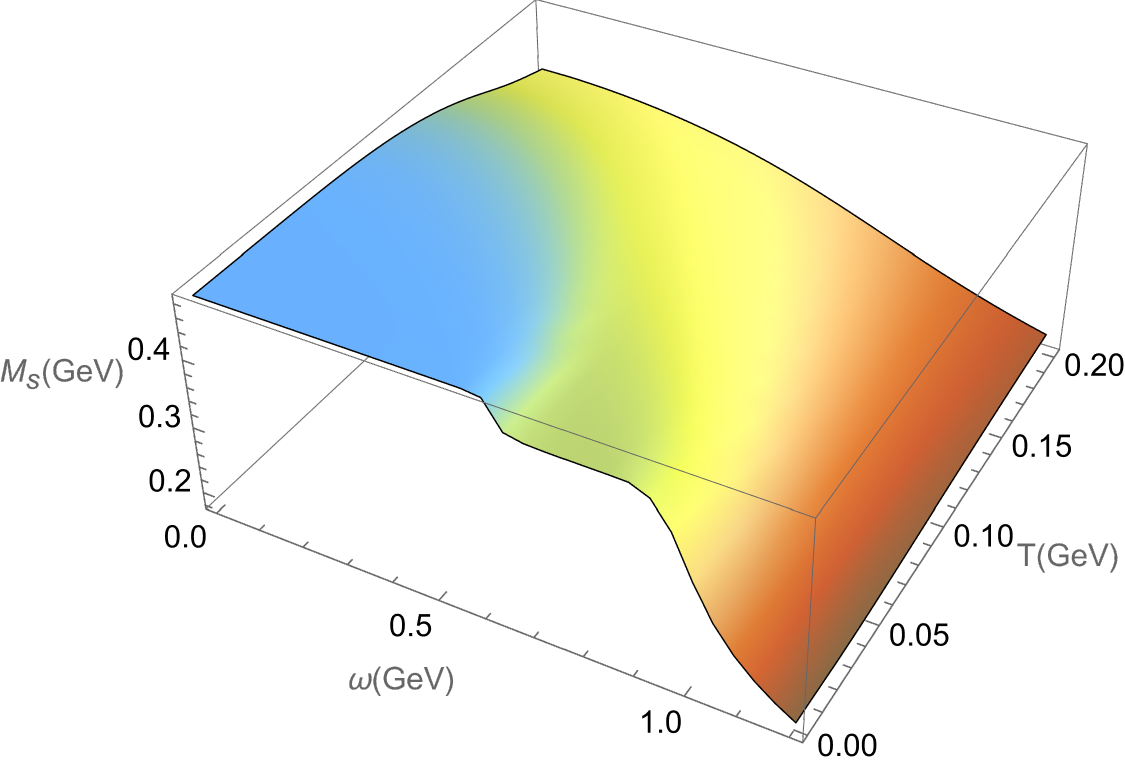}}
\caption[]{(Color online) The effective mass of light quark and strange quark according to  temperature $T$ and angular velocity $\omega$ with $\mu=0.01$ GeV.}
\label{MtoTOmega.pdf}
\end{figure}

We present the evolution of the light quark mass with respect to $T$, $\omega$  in \Cref{MtoTOmega.pdf}(a), and the strange quark mass  in \Cref{MtoTOmega.pdf}(b). We observe a decrease in mass as the temperature or angular velocity increases, indicating the restoration of chiral symmetry at high temperatures or large angular velocities. It is remarkable that there exist fast transitions in the low temperature and large angular velocity region, while in the high temperature and small angular velocity region only exhibits a very slow change. It is easy to see that when we fix the temperature is very low, the restoration of chiral symmetry experiences a fast transition with rapid rotation. This can be seen as the continuous crossover becoming steeper with increasing angular velocity, eventually merging into a fast transition at large angular velocity. By comparing the critical angular velocities $\omega_{c}$ for the light and strange quarks, we find that the decrease is faster for the light quark, which suggests that the chiral symmetry restoration is more efficient for the light quark compared to the strange quark.
\begin{figure}
\subfigure[]
{\includegraphics[width=0.45\textwidth]{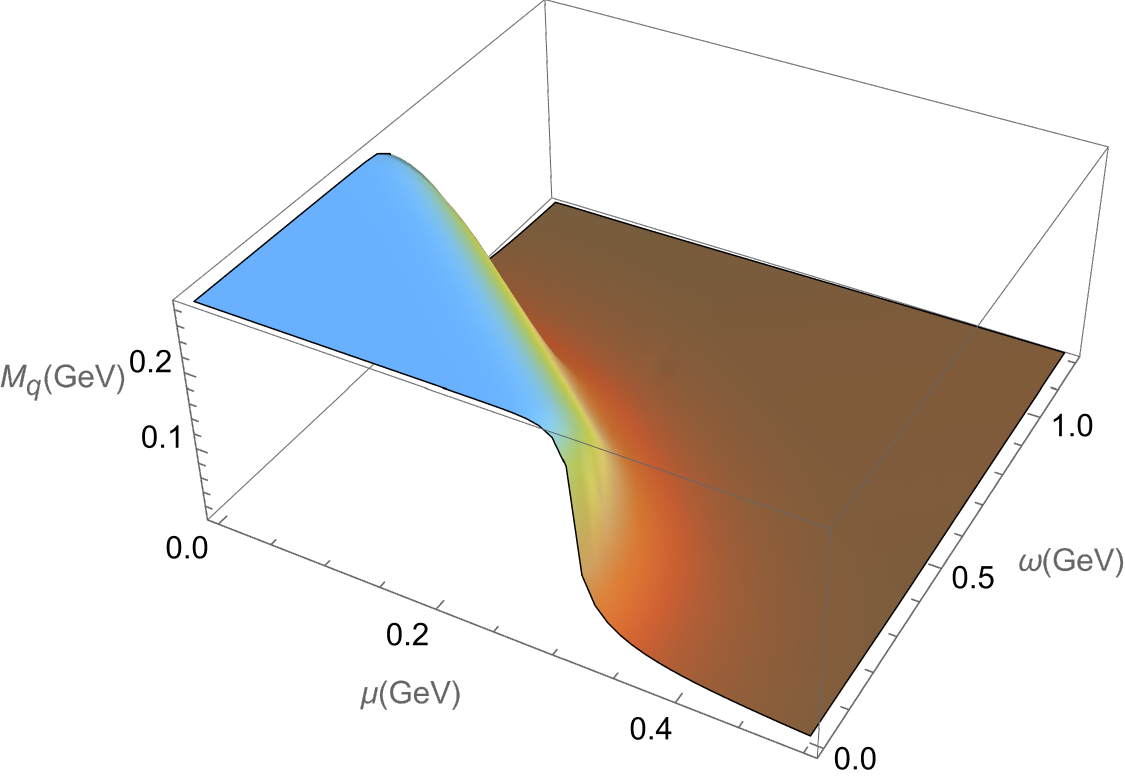}}
\subfigure[]
{\includegraphics[width=0.45\textwidth]{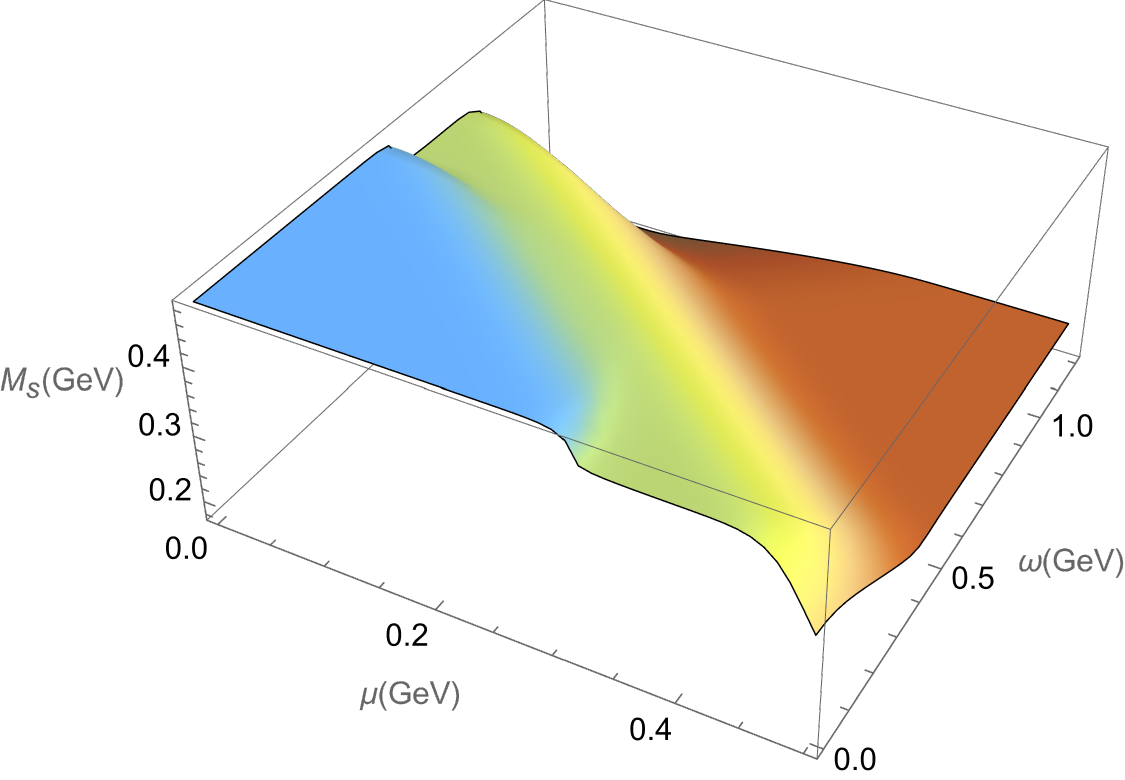}}
\caption[]{(Color online) The effective mass of light quark and strange quark according to angular velocity $\omega$ and quark chemical potential $\mu$ with $T=0.01$ GeV.}
\label{MtoMuOmega.pdf}
\end{figure}
\begin{figure}
\subfigure[]
{\includegraphics[width=0.45\textwidth]{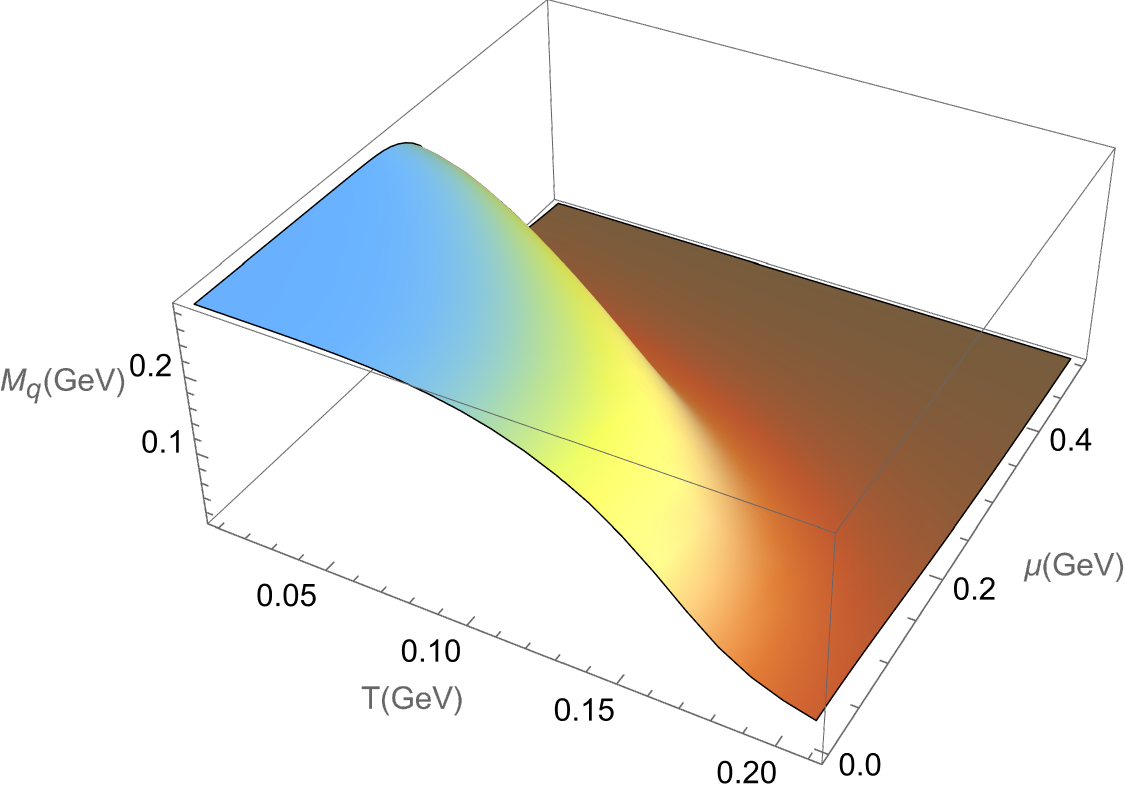}}
\subfigure[]
{\includegraphics[width=0.45\textwidth]{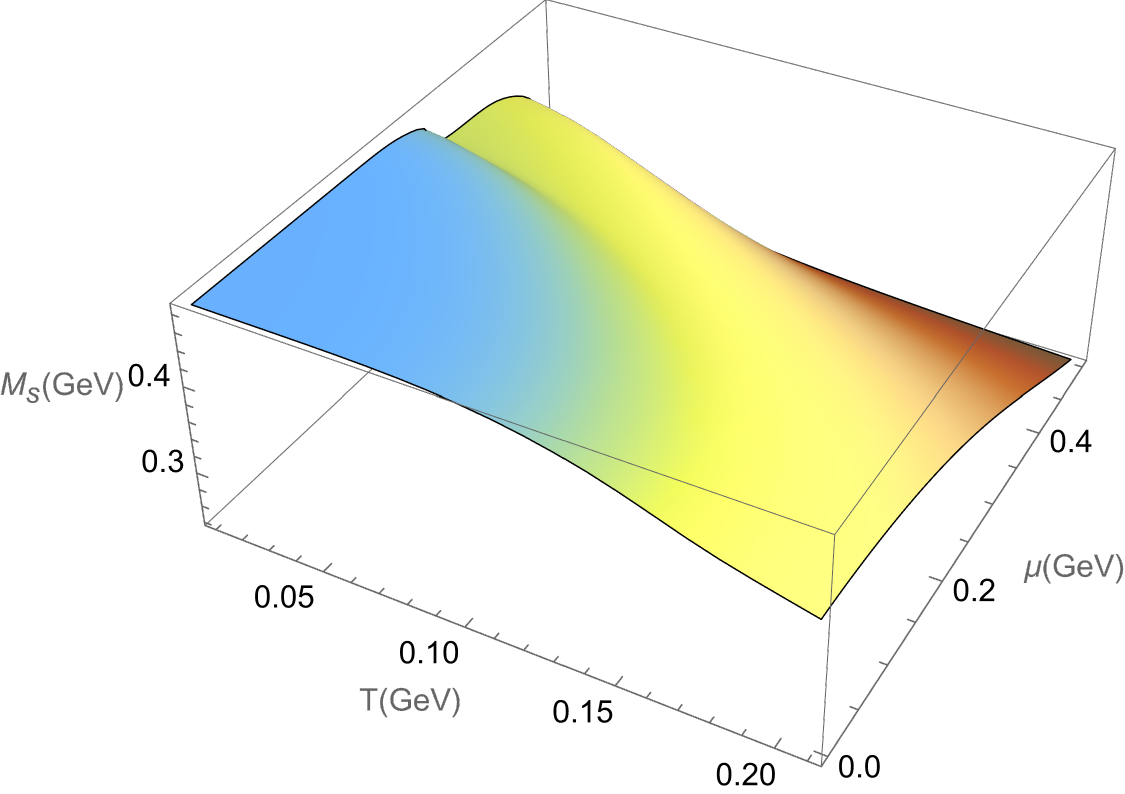}}
\caption[]{(Color online) The effective mass of light quark and strange quark according to temperature $T$ and quark chemical potential $\mu$ with $\omega=0.1$ GeV.}
\label{MtoTMu.pdf}
\end{figure}
\begin{figure}
\subfigure[]
{\includegraphics[width=0.45\textwidth]{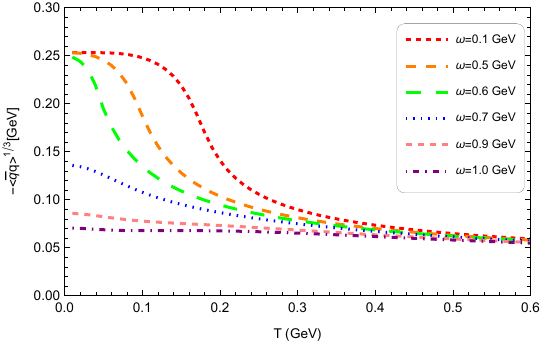}}
\subfigure[]
{\includegraphics[width=0.45\textwidth]{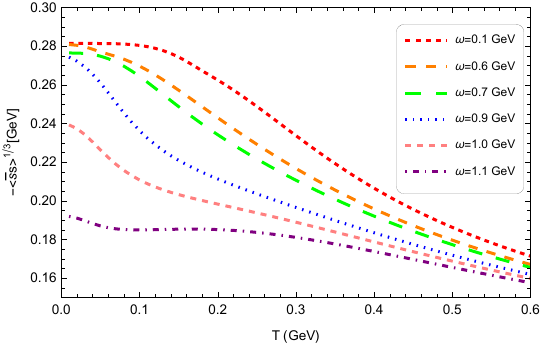}}
\caption[]{(Color online) Condensates of light quark  and strange quark as functions of temperature for different values of the rotational speed.}
\label{condensate.pdf}
\end{figure}
\begin{figure}
{\includegraphics[width=0.45\textwidth]{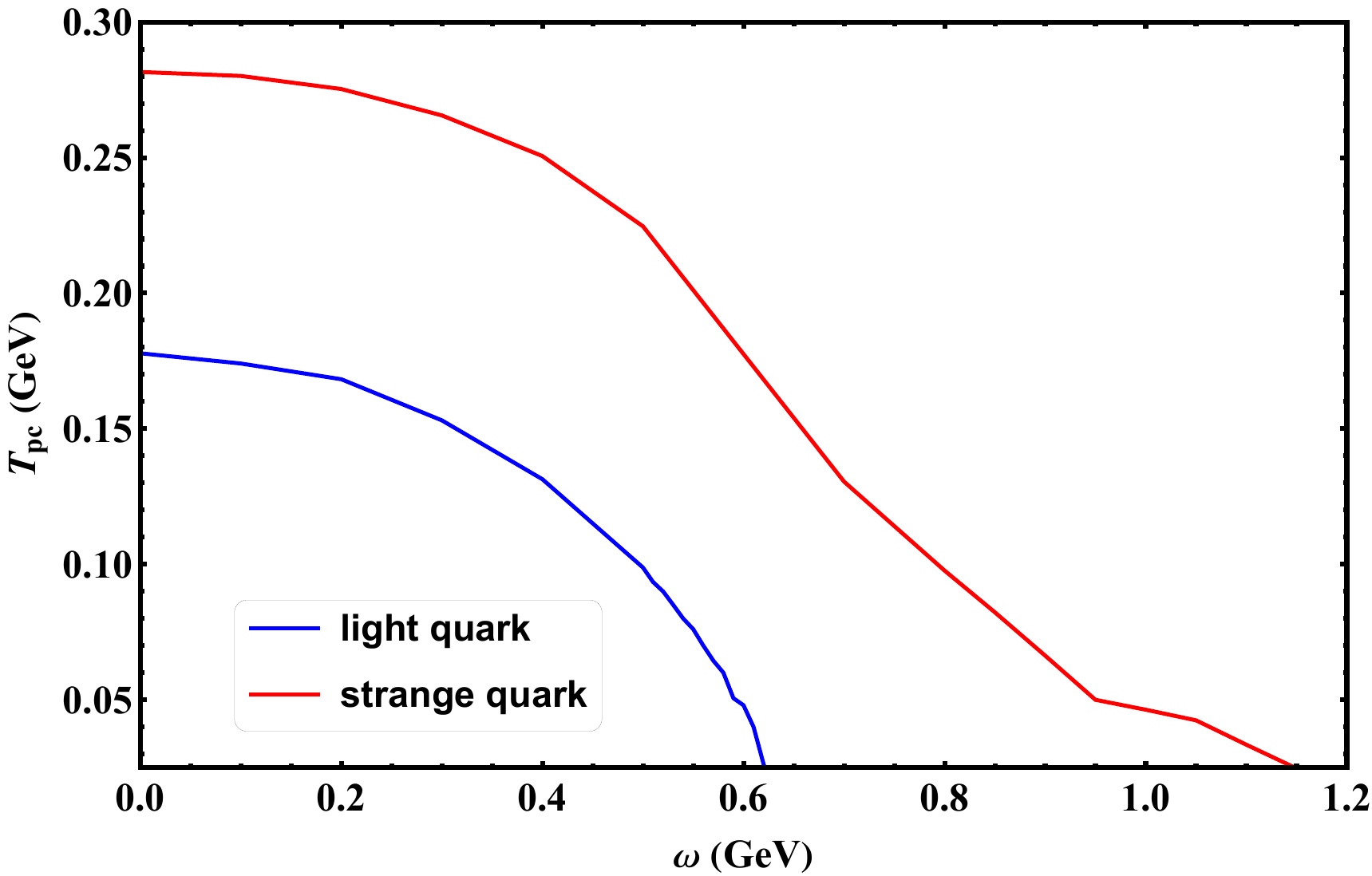}}
\caption[]{(Color online) The pseudocritical temperatures for the chiral transition of rotating quark matter as functions of the angular velocity.}
\label{transition.pdf}
\end{figure}

Then, we extended the investigation of effective quark mass to the $\omega-\mu$  plane. In \Cref{MtoMuOmega.pdf}, we show the evolution of the effective masses for the light quark and strange quark as functions of the angular velocity and quark chemical potential. Clearly, at sufficiently large angular velocity or (and) sufficiently large quark chemical potential the quark effective mass is very small. When angular velocity is small, we can find there is a very swift change for the light quark around $\mu=0.3$ GeV, and for the strange effective quark there are two quickly change regions. We can also observe similar transition regions of small quark chemical potential and large angular velocity as shown in the left rear side of the figure.

The temperature and chemical potential dependence of the light and strange quark effective masses at $\omega=0.1$ GeV is depicted in \Cref{MtoTMu.pdf}, one sees that at low temperature and small quark chemical potential, the chiral symmetry is spontaneously broken, with increasing temperature or quark chemical potential, the effective mass of strange quark is only mildly dependent on them while the light quark shows more sensitive to them compared the strange quark. It can be found that in the low temperature region, the effective mass of the light quark has a sharp drop at certain values of $\mu$ with increasing $\mu$. One can see at high temperature or large quark chemical potential the effective mass of these quarks become small, and at sufficiently high temperature or sufficiently large chemical potential the effective mass of light quark almost approaches its current mass, while the strange quark still with a heavy effective mass, this indicates that the current mass of the quark plays an important role in the chiral transition.

The quark condensate $\left\langle {\bar qq} \right\rangle$ or $\left\langle {\bar ss} \right\rangle$ is often treated as an order parameter for spontaneous chiral symmetry breaking. The temperature-dependence of the order parameters for different values of the rotational speed are shown in \Cref{condensate.pdf}. It shows a rapid cross-over with a critical angular velocity at about $0.6$ GeV and $1.0$ GeV for the light quark and strange quark condensate, respectively. At low temperatures and small angular velocities, the chiral symmetry is spontaneously broken. While at high temperatures or (and) large angular velocities, chiral symmetry may gradually be restored.

Next, we determine the chiral phase transition temperature in the presence of angular velocity in \Cref{transition.pdf}. The definition of $T_{pc}$ in this context is determined by the maximum of $\left| {\frac{{d{\phi _f}}}{{dT}}} \right|$, here, $f=u,d,s$ and ${\phi _u} ={\phi _d} = \left\langle {\bar qq} \right\rangle ,{\phi _s} = \left\langle {\bar ss} \right\rangle$. From \Cref{transition.pdf} we can see the pseudocritical temperature decreases as angular velocity becomes larger. At small angular velocity region, the pseudocritical temperature of strange quark is about $0.1$ GeV larger than that of the light quark, even at a large angular velocity around $0.6$ GeV, where the pseudocritical temperature of light quark is very small and by contrast the pseudocritical temperature of strange is still very large. Thus, a conclusion seems to be that the rotation can lead to an obvious change to the chiral transition of light quarks compared to that of strange quark due to whose mass is heavier.

\subsection{Thermodynamics results for different angular velocities at vanishing quark chemical potential}

\begin{figure}
\subfigure[]
{\includegraphics[width=0.4\textwidth]{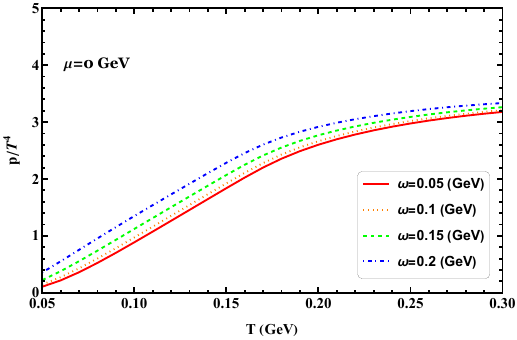}}
\subfigure[]
{\includegraphics[width=0.4\textwidth]{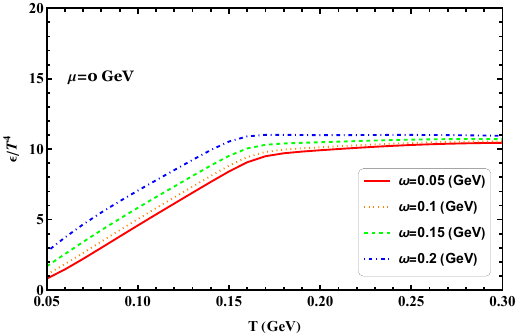}}
\subfigure[]
{\includegraphics[width=0.4\textwidth]{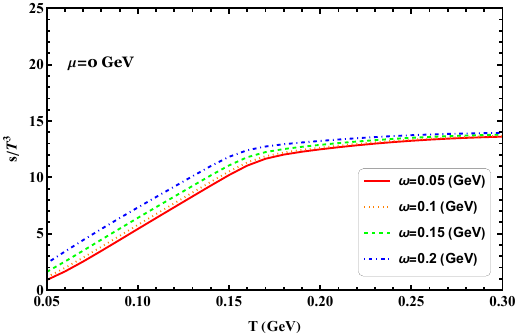}}
\subfigure[]
{\includegraphics[width=0.4\textwidth]{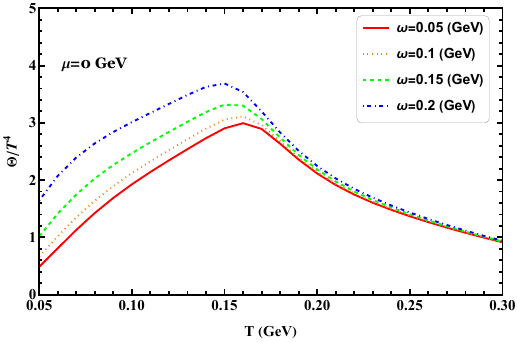}}
\caption[]{(Color online) Scaled pressure, energy density, entropy density and trace anomaly as functions of temperature at zero chemical potential for different angular velocities.}
\label{Scaledpeetmu0.pdf}
\end{figure}
\begin{figure}
\subfigure[]
{\includegraphics[width=0.45\textwidth]{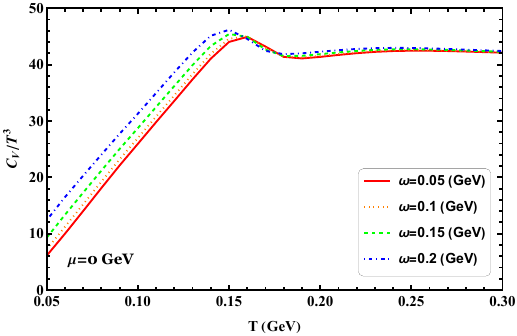}}
\subfigure[]
{\includegraphics[width=0.45\textwidth]{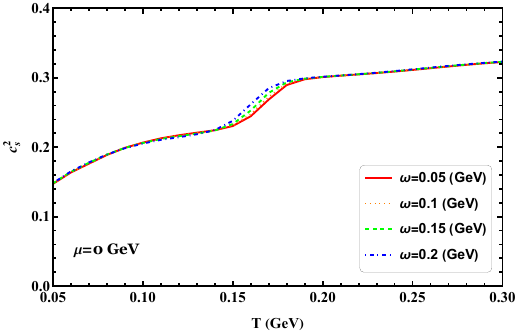}}
\caption[]{(Color online) Scaled specific heat and speed of sound squared as functions of temperature at zero chemical potential for different angular velocities.}
\label{cvcsmu0.pdf}
\end{figure}
\begin{figure}
\subfigure[]
{\includegraphics[width=0.45\textwidth]{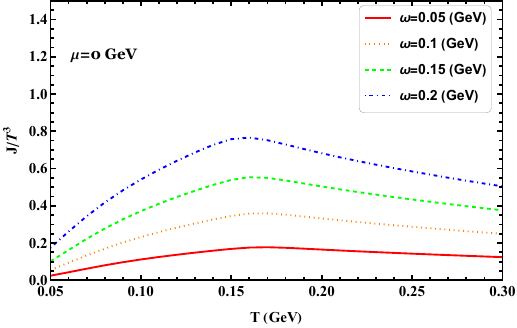}}
\subfigure[]
{\includegraphics[width=0.45\textwidth]{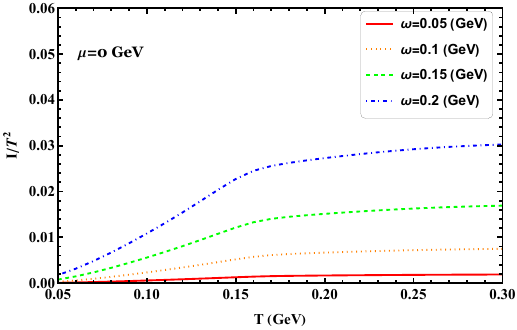}}
\caption[]{(Color online) Scaled angular momentum and moment of inertia as functions of temperature at zero chemical potential for different angular velocities.}
\label{ammi.pdf}
\end{figure}
As can be seen in \Cref{Scaledpeetmu0.pdf}, the scaled pressure, energy, and entropy densities increase with increasing temperature. These quantities start to increase rapidly with temperature and then gradually grow as the temperature continues to rise after passing through the transition region. The observable rotation enhances these scaled quantities, at low temperatures, the enhancements are significant, while at high temperatures, the enhancements are less pronounced. The scaled trace anomaly exhibits a peak in the transition region, as the temperature continues to increase, the scaled trace anomaly decreases for all different angular velocities. It is evident that at low temperatures the variation in angular velocity can result in significant deviations of the scaled trace anomaly, and this effect diminishes as the firing temperature increases.

The specific heat is an important quantity in thermodynamics as it can be considered a response function of the phase transition, its variation with temperature is presented in \Cref{cvcsmu0.pdf}(a). As the angular velocity increases, the peak of the specific heat, which occurs at the transition temperature, shifts towards lower temperatures, this indicates that the transition temperature decreases with an increase in angular velocity. In \Cref{cvcsmu0.pdf}(b), the speed of sound squared increases with temperature and shows little sensitivity to the chosen angular velocities. However, this observation may be attributed to the consideration of only small values of angular velocity. It is evident that the speed of sound squared approaches the conformal limit of 1/3 for different angular velocities at high temperature limits.

An intriguing quantity in the rotating system is the angular momentum. \Cref{ammi.pdf}(a) displays the results of the scaled angular momentum as functions of temperature at zero chemical potential for various angular velocities. The scaled angular momentum initially increases with temperature and reaches its peak across the chiral transition region ($\sim 150$ MeV) for all angular velocities. Beyond this temperature, it decreases with further temperature increment. The moment of inertia is also of interest in our calculation as it represents the linear response of the system's angular momentum $J$ to the angular velocity $\omega$. \Cref{ammi.pdf}(b) displays the results of the  moment of inertia as functions of temperature at zero chemical potential for various angular velocities. It is evident that the scaled moment of inertia always increases with temperature for different angular velocities. Moreover, for a fixed temperature, the scaled moment of inertia becomes larger with increasing angular velocity.\\

\subsection{The influence of the radius on the thermodynamics in the rotating system}
\begin{figure}
\subfigure[]
{\includegraphics[width=0.4\textwidth]{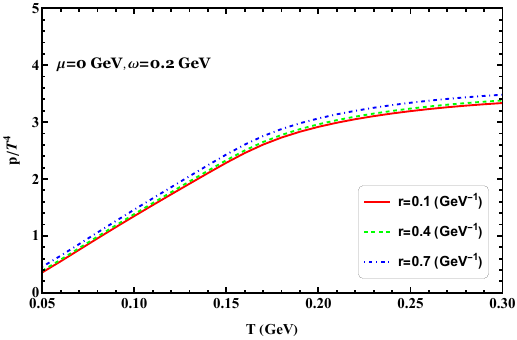}}
\subfigure[]
{\includegraphics[width=0.4\textwidth]{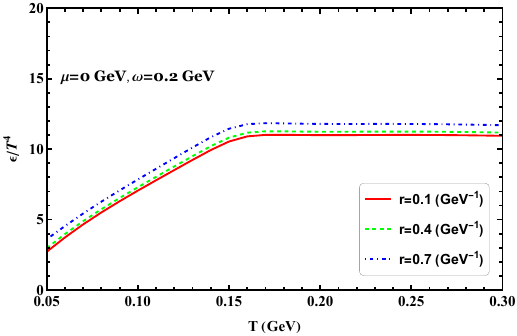}}
\subfigure[]
{\includegraphics[width=0.4\textwidth]{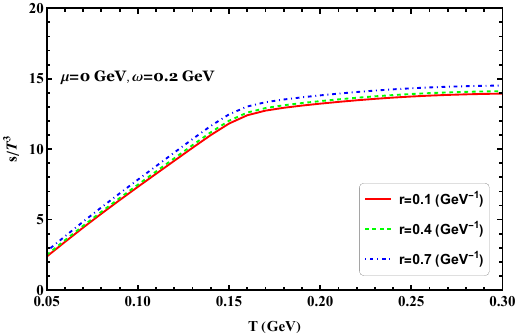}}
\subfigure[]
{\includegraphics[width=0.4\textwidth]{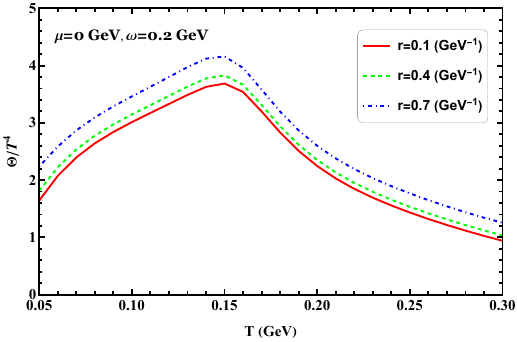}}
\caption[]{(Color online) Scaled pressure, energy density, entropy density and trace anomaly as functions of temperature at zero chemical potential for different radii.}
\label{Scaledpeetr.pdf}
\end{figure}
\begin{figure}
\subfigure[]
{\includegraphics[width=0.45\textwidth]{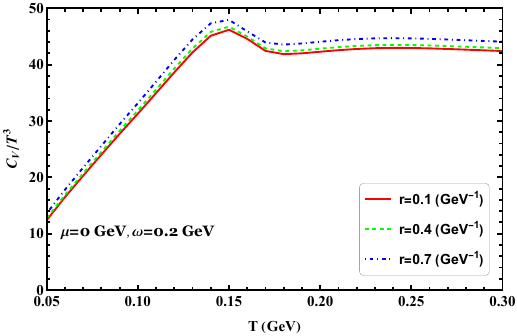}}
\subfigure[]
{\includegraphics[width=0.45\textwidth]{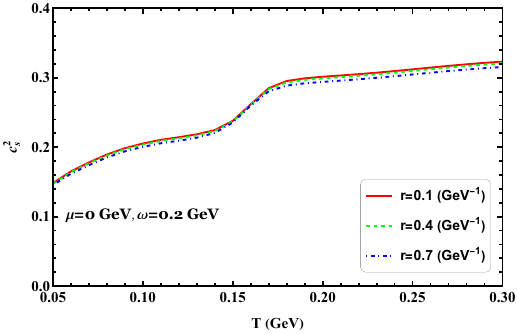}}
\caption[]{(Color online) Scaled specific heat and speed of sound squared as functions of temperature at zero chemical potential for different radii.}
\label{cvcsmu0r.pdf}
\end{figure}

In the directions perpendicular to the rotating axis, the rotating radius should be a finite value determined by the causal condition $\omega r<1$. It has been expected that the presence of boundaries can modify the properties of a rotating system \cite{Chen:2015hfc,Ebihara:2016fwa,Chernodub:2016kxh,Chernodub:2017ref,Chernodub:2020qah}. However, this is only true when the angular velocity $\omega$ is much smaller than the inverse of the system's size \cite{Wang:2018sur}. In the present calculations, we will neglect the finite volume effect and consider it in future studies.

In the standard NJL model, these thermodynamics are functions of temperature and quark chemical potential. However, in a rotating system, these thermodynamics should also depend on the finite size. Due to the cylindrical symmetry, these quantities are dependent on the transverse radius $r$. It would be interesting to investigate how the various thermodynamic quantities in a strongly interacting rotating matter depend on the radius of the rotating system. The properties as a function of the radius of the rotating system may be related to experimental observations in the future. Additionally, the radius should drastically change the angular momentum and the momentum of the inertia. Therefore, it becomes important to study how the various thermodynamic quantities in the QCD matter under rotation depend on the rotation radius of the system.

We show the  densities of the scaled pressure, energy, entropy, and trace anomaly as functions of temperature at zero chemical potential for different radii in \Cref{Scaledpeetr.pdf}. As can be seen, the radius effect is visible and enhances these thermodynamic quantities. The radius effect does not qualitatively affect the behavior of these thermodynamic quantities even in the high-temperature region; it just shifts these thermodynamics at a given temperature. It is also noted that the differences for each thermodynamic quantity between different radii seem unchanged even at high-temperature region.

Scaled angular momentum and moment of inertia as functions of temperature at zero chemical potential for different radii are shown in \Cref{cvcsmu0r.pdf}. In the \Cref{cvcsmu0r.pdf}(a) we find that there is a rapid change near the chiral transition region (around 150 MeV) for the scaled specific heat. It can be seen that there exists a characteristic where the location of the summits almost to have no change with increasing radius. In \Cref{cvcsmu0r.pdf}(b), it is remarkable that the speed of sound squared curves seem the same around the chiral transition region. It is also found that at extremely high temperatures, for different angular velocities, all the values of the speed of sound squared approach the Stefan-Boltzmann limit, whose value is 1/3. This indicates that the speed of sound squared of quark matter at high temperature is not sensitive to the transverse radius.

\begin{figure}
\subfigure[]
{\includegraphics[width=0.45\textwidth]{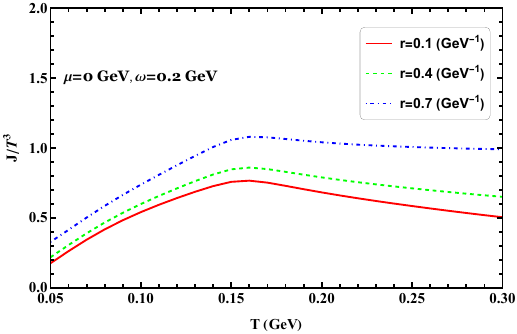}}
\subfigure[]
{\includegraphics[width=0.45\textwidth]{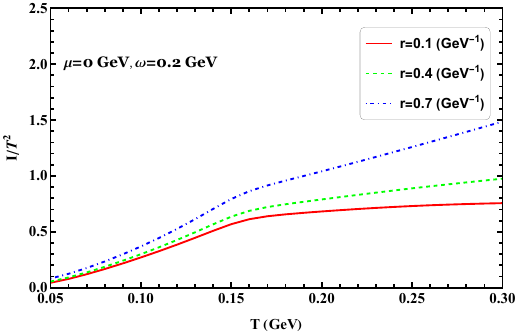}}
\caption[]{(Color online) Scaled angular momentum and moment of inertia as functions of temperature at zero chemical potential for different radii.}
\label{ammir.pdf}
\end{figure}

From \Cref{ammir.pdf}(a), one could also infer the dependence of the scaled angular momentum on the radius. Unlike other thermodynamics, the angular momentum has a strong dependence on the system radius. At low temperatures, the angular momentum increases smoothly with increasing radius. At high temperatures, the angular momentum becomes stronger. It is also evident that the scaled moment of inertia always increases with increasing temperature for different radii from \Cref{ammir.pdf}(b). In the region of high temperature, the scaled moment of inertia shows a strong radius dependence. As the temperature increases, the difference between any two curves in the figure becomes larger for both quantities.

\subsection{Thermodynamics in rotating system at finite chemical potential}

\begin{figure}
\subfigure[]
{\includegraphics[width=0.42\textwidth]{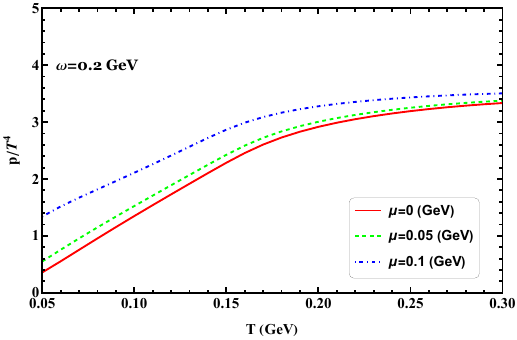}}
\subfigure[]
{\includegraphics[width=0.42\textwidth]{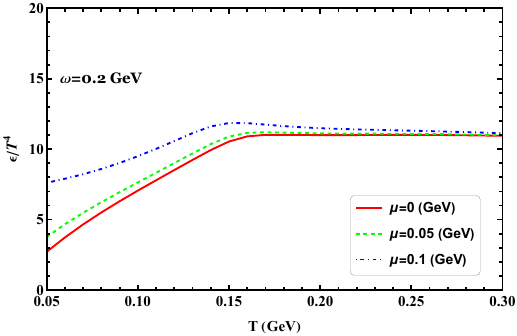}}
\subfigure[]
{\includegraphics[width=0.42\textwidth]{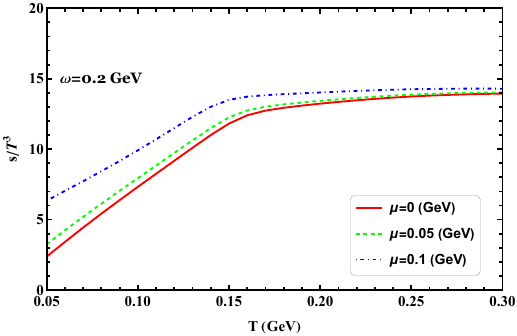}}
\subfigure[]
{\includegraphics[width=0.42\textwidth]{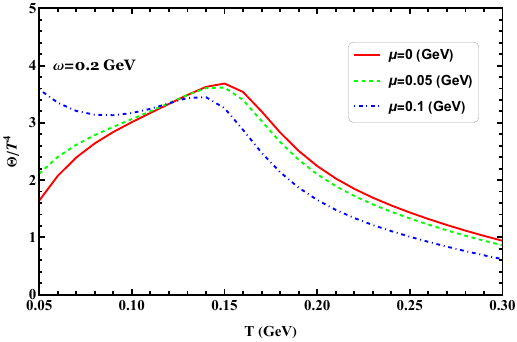}}
\caption[]{(Color online) Scaled pressure, energy density, entropy density and trace anomaly as functions of temperature at $\omega=0.2$ GeV for different chemical potentials.}
\label{Scaledpeetmu.pdf}
\end{figure}
Studying the thermodynamics at finite chemical potential in the rotating system is important for understanding the phase structure of QCD, modeling compact stars, and interpreting heavy ion collision experiments. In \Cref{Scaledpeetmu.pdf} we show the densities of scaled pressure, energy, entropy, and trace anomaly as functions of temperature at $\omega=0.2$ GeV for different chemical potentials. It can be easy to see that, at low temperature region, there can be a nontrivial contribution from the chemical potential. It is shown that the pressure, energy, and entropy density increase with increasing for different angular velocities, and these quantities are also enhanced by the chemical potential. An increase in the chemical potential leads to increases in these thermodynamics, which can be easily understood as more degrees of freedom are active. In the rotating system, the trace anomaly is enhanced by the chemical potential below the critical transition region, while across the transition region, it can be found the trace anomaly is suppressed by the chemical potential, in addition, with increasing chemical potential, the crossover pattern evolves to lower transition temperatures. For all the quantities in this figure show that below the crossover temperature they exhibit a strong chemical potential dependence.
\begin{figure}
\subfigure[]
{\includegraphics[width=0.45\textwidth]{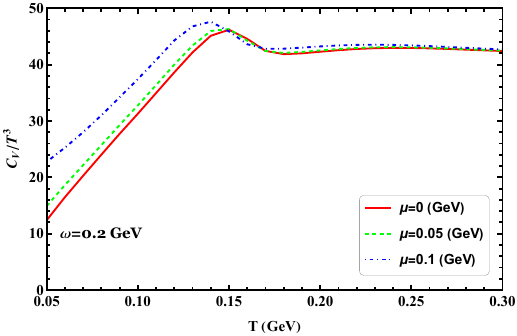}}
\subfigure[]
{\includegraphics[width=0.45\textwidth]{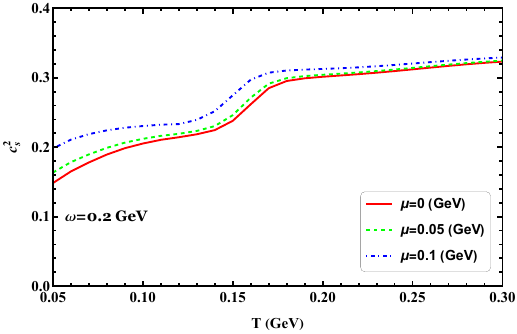}}
\caption[]{(Color online) Scaled specific heat and speed of sound squared as functions of temperature at $\omega=0.2$ GeV for different chemical potential.}
\label{cvcsmu.pdf}
\end{figure}
In \Cref{cvcsmu.pdf} we show scaled specific heat and speed of sound squared as functions of temperature at $\omega=0.2$ GeV for different chemical potentials. As shown in \Cref{cvcsmu.pdf}(a), scaled specific heat increases with increasing temperature and reaches a peak at the chiral transition region. Then it first decreases quickly around the critical chiral transition and finally changes little with temperature. The figure shows that the peak position moves to a smaller temperature as quark chemical potential increases. From \Cref{cvcsmu.pdf}(b), we can see there is a significant increase in the speed of sound squared for increased quark chemical potential, even near the transition region, which means in the finite chemical potential may have an important effect on the thermalization of the QCD matter in the rotating system. Here, the speed of sound squared also conveys relevant information: it displays no local minimum at a crossover transition for the quark chemical potential considered, due to first is that the system is not an infinite volume as the standard NJL model, another reason is that the energy density has been modified, ie. we add the contribution of $J\omega$. It is not hard to see there is a trend when increasing chemical potential in the rotating system, there will appear a local minimum in the phase transition. From this figure, we can see the speed of sound squared will approach the conformal limit of 1/3 for different angular velocities at a large temperature limit.

\begin{figure}
\subfigure[]
{\includegraphics[width=0.45\textwidth]{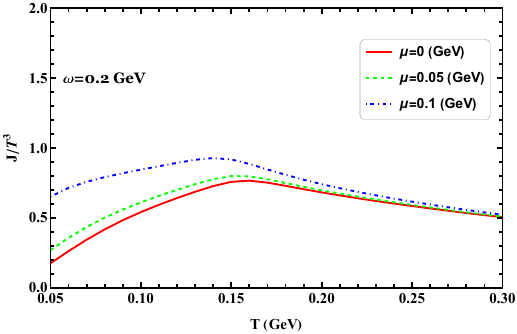}}
\subfigure[]
{\includegraphics[width=0.45\textwidth]{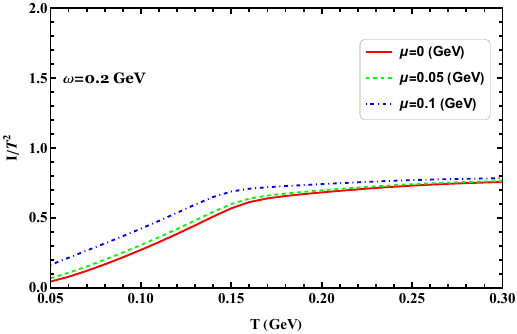}}
\caption[]{(Color online) Scaled angular momentum and moment of inertia as functions of temperature at $\omega=0.2$ GeV for different chemical potential.}
\label{ammirmu.pdf}
\end{figure}

Another basic thermodynamic quantity is the angular momentum and moment of inertia, these quantities measure the breaking of conformal symmetry in the interaction theory. In \Cref{ammirmu.pdf}, we show the scaled angular momentum and moment of inertia as functions of the temperature for different chemical potential at finite angular velocities. They have similar characteristics as scaled quantities in \Cref{Scaledpeetmu.pdf} below the critical transition. When continue increasing the temperature, the scaled angular momentum slowly decreases, while the scaled moment of inertia always keeping increasing with temperature.
\begin{figure}
{\includegraphics[width=0.45\textwidth]{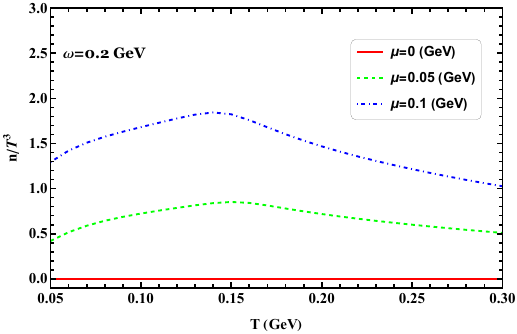}}
\caption[]{(Color online) Scaled quark number density as a function of temperature at $\omega=0.2$ GeV for different values of chemical potential.}
\label{quarknumberaddchemical.pdf}
\end{figure}

Another possible signature of the chiral transition is offered by the behavior of the quark number densities.
In \Cref{quarknumberaddchemical.pdf} we show the results of the scaled quark number density as a function of the temperature at $\omega=0.2$ GeV for different values of chemical potential, from the figure we can see when quark chemical potential equals to zero, the corresponding quark number density is always zero. In the presence of finite quark chemical potential, the scaled quark number densities increase slightly until at $T=150$ GeV and decrease again with growing temperature. It is obvious that the chemical potential enhances the quark number density in the rotation system.

\subsection{Thermodynamics in rotating system at large angular velocity}
\begin{figure}
\subfigure[]
{\includegraphics[width=0.42\textwidth]{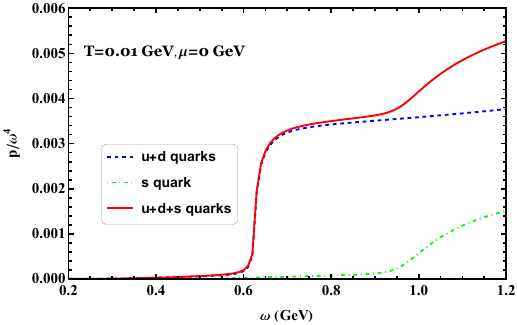}}
\subfigure[]
{\includegraphics[width=0.42\textwidth]{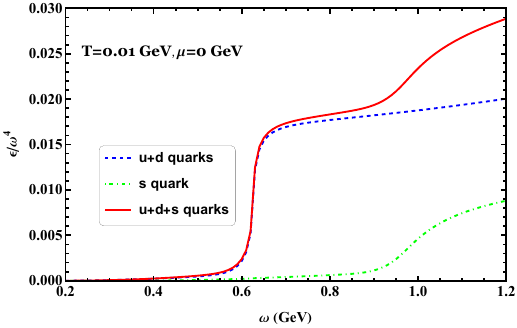}}
\subfigure[]
{\includegraphics[width=0.42\textwidth]{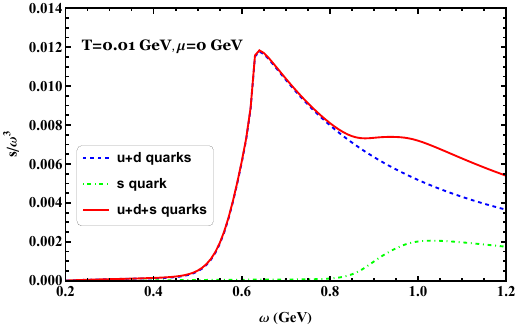}}
\subfigure[]
{\includegraphics[width=0.42\textwidth]{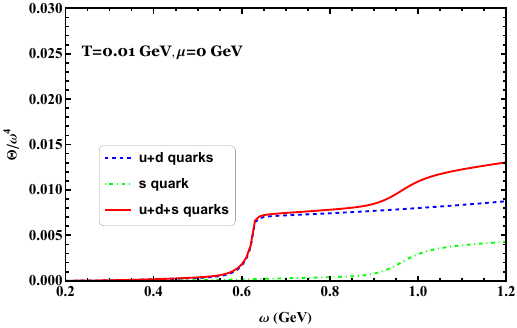}}
\caption[]{(Color online) Scaled pressure, energy density, entropy density and trace anomaly as functions of angular velocity at $T=0.01$ GeV and $\mu=0$ GeV for the light, strange and total quarks, respectively.}
\label{Scaledpeetomega.pdf}
\end{figure}
\begin{figure}
{\includegraphics[width=0.45\textwidth]{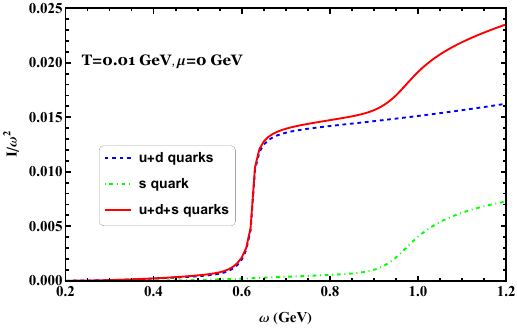}}
\caption[]{(Color online) Scaled moment of
inertia as functions of angular velocity at $T=0.01$ GeV and $\mu=0$ GeV for the light, strange and total quarks, respectively.}
\label{ammiromega.pdf}
\end{figure}

In the following, we will present a systematic analysis of the thermodynamic quantities of QCD matter under large angular velocity. The system's total pressure and energy density during rotation are simply the sum of the contributions from each quark flavor. In order to have a clearer picture of the effects of rotation on different quark flavors, we will investigate each individual contribution as well as the total contribution.

From the strong rotational behavior depicted in \Cref{Scaledpeetomega.pdf}, it is evident that the bulk thermodynamic properties, such as the scaled pressure, energy density, and trace anomaly, increase with increasing angular velocity at a temperature of  $T=0.01$ GeV and quark chemical potential $\mu=0$ GeV. Notably, both the scaled pressure, energy, and trace anomaly of the light quark and strange quark exhibit an increase as the angular velocity rises. In the mid-region of angular velocity, below approximately $0.8$ GeV, the light quark predominantly contributes to these thermodynamic quantities. However, at sufficiently large angular velocities, the contributions from different flavors become almost the same. It can be also found that the angular momentum of the system has also a very similar character in \Cref{ammiromega.pdf}. It is evident that the angular momentum in the chiral broken phase is lower than the angular momentum in the chiral restored phase. Furthermore, it is worth noting that the contribution of the light quark to the angular momentum is remarkable in the mid-region of the angular velocity, while that of the strange quark is moderate.

There is a descent for the scaled entropy density after exceeding the critical point around $\omega=0.6$ GeV because, in this region, the rate of increase of the quantity pressure is slowing down. We can also observe a slight increase (not clearly visible in the figure) followed by a decrease in the entropy density around $\omega=1.0$ GeV. The trace anomaly increases with increasing angular velocity, which is because we have set $T=0.01$ GeV, in such low temperature, the strange quark is still in a phase with partly broken chiral symmetry if the temperature is high, we will see that the trace anomaly becomes small.
\begin{figure}
\subfigure[]
{\includegraphics[width=0.45\textwidth]{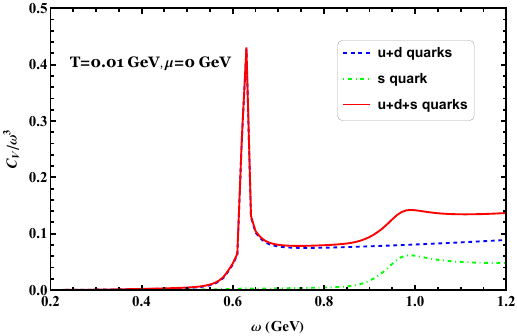}}
\subfigure[]
{\includegraphics[width=0.45\textwidth]{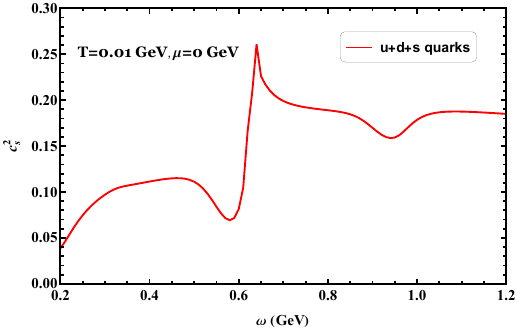}}
\caption[]{(Color online) (a) Scaled specific heat as functions of angular velocity at $T=0.01$ GeV and $\mu=0$ GeV for the light, strange and total quarks, respectively. (b) The corresponding result of speed of sound squared for the total quarks.}
\label{shssomega.pdf}
\end{figure}
\begin{figure}
{\includegraphics[width=0.45\textwidth]{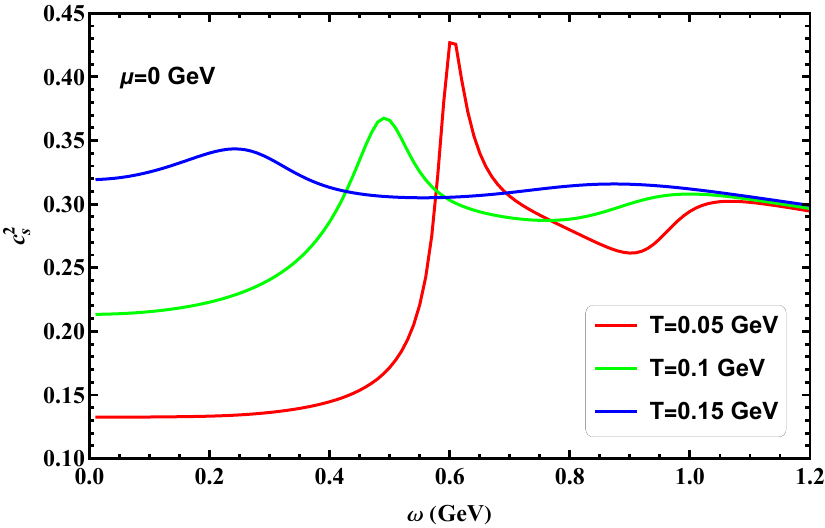}}
\caption[]{(Color online) Speed of sound squared as function of angular velocity at $\mu=0$ GeV for different temperatures.}
\label{speedofsoundomegaT.pdf}
\end{figure}

We show the behavior of the scaled specific heat as a function of $\omega$ in \Cref{shssomega.pdf}(a) at $T=0.01$ GeV for vanishing chemical potential. The evolution of the scaled specific heat increases from zero to a maximum value around $\omega=0.6$ GeV then down to a minimum value and then gradually increases to another sub-maximum value around $\omega=1.0$ GeV and finally tends to gradually decrease at large angular velocity. From the figure, we can clearly see that the light quark rises steeply across the chiral transition and for the strange quark only there is a flatter peak at a more relatively broad region. It is known that if one has a sharp crossover phenomenon with a rapid change in thermodynamic quantities over a small interval, there is some chance for measurable effects in experiments, so the specific heat may provide relevant signatures for phase transitions in the rotating system.

The speed of sound squared changes with the angular velocity for the light and strange quarks is plotted in \Cref{shssomega.pdf}(b). It is known that in the transition region of the QCD matter, the characteristics of the speed of sound squared undergo significant changes, it is evident that the speed of sound squared shows a (pronounced) dip near the chiral transition. There are two local minimums of the speed of sound squared becomes deeper in the vicinity of the critical angular velocity, which correspondingly to the light quark and strange quark, respectively. At small angular velocity, the speed of sound squared increases with an increase in the angular velocity. However, at a large angular velocity, the speed of sound squared subtly decreases with an increase in the angular velocity. Our numerical results indicate that the dependence of the speed of sound squared on the angular velocity can be indicative of QCD chiral transition. To probe this dependence further, we show the results of calculations for different temperatures in \Cref{speedofsoundomegaT.pdf}, the figure exhibits markedly behavior of the quark matter under rotation, and it can be found that the maximum value of the speed of sound is dominated by features associated with the chiral transitions. In addition, the speed of sound squared increases as angular velocity increases in the small angular velocity region, while decreases with angular velocity increases in the large angular velocity region.  In the low angular velocity region, there is a significant difference in the speed of sound corresponding to different temperatures. However, in the high angular velocity region, the difference in the speed of sound becomes smaller for different temperatures and ultimately converges to the same value. Thus, a key conclusion can be made that the speed of sound exhibits a nonmonotonic behavior as the angular velocity changes.

\section{CONCLUSIONS\label{sec4}}

In order to investigate the expansion of the plasma formed in ultra-relativistic heavy-ion collisions with noncentral impact, it is crucial to compute the thermodynamic properties within a rotating system. This paper focuses on formulating and exploring the thermodynamics of the three-flavor NJL model under rotation. We present the outcomes concerning diverse thermodynamic observables as a function of temperature, considering various angular velocities, radii, and finite quark chemical potentials. Additionally, we examine the thermodynamic behaviors of the light and strange quarks in relation to the angular velocities, respectively.

To summarize, we have presented an analytical calculation of the thermodynamics in the three-flavor NJL model in the presence of the rotational effect. We systematically analyze the equation of state in the parameter space of temperature $T$, chemical potential $\mu$, and the angular velocity $\omega$ in the rotational system. The calculations provide a physical picture of the chiral transition under rotation, and our findings indicate that the effect of rotation plays an important role in thermodynamics.  By studying the changes in thermodynamic quantities of a rotating system, we can gain insights into the properties and behaviors of QCD matter. In the rotating system, the scaled thermodynamic quantities are visibly influenced by rotation, an important quantity is the moment of inertia, which exhibits a strong dependence on the angular velocity even at high temperatures. The thermodynamic properties of light and heavy quarks differ with respect to different angular velocities, and this distinction strongly influences the thermodynamic quantities, however, for sufficiently strong rotation, these distinctions for each flavor vanish.  The speed of sound plays a crucial role in studying the thermodynamic properties and phase transitions of QGP, as a main finding of our analysis, the speed of sound squared exhibits a nonmonotonic feature with respect to the angular velocity.

It is important to mention that here, for simplicity, we did not take into account the boundary effect of the system. Since any uniformly rotating system should be spatially bounded, it has been expected that the presence of boundaries can modify the properties of the rotating system \cite{Chen:2015hfc,Ebihara:2016fwa,Chernodub:2016kxh,Chernodub:2017ref,Chernodub:2020qah}, indeed, this is only true when the angular velocity is much smaller than the inverse of the system's size
\cite{Wang:2018sur}, so in our analytic derivation we
ignore the finite volume boundary effect and we leave it as our further study. So far, we have developed the NJL model taking into account only fermion-antifermion scalar interactions for the chiral transition, it is necessary to note that the vector interactions  \cite{Asakawa:1989bq,Klimt:1990ws,Buballa:1996tm} may play an important role on the chiral transition of three-flavor NJL model in the present of rotation, and we also leave it as our further study. In addition, the Polyakov-Nambu-Jona-Lasinio (PNJL) model \cite{Meisinger:1995ih,Meisinger:2001cq,Fukushima:2003fw,Mocsy:2003qw,Megias:2004hj,Ratti:2005jh,Fukushima:2008wg} incorporates the Polyakov loop integral based on the NJL model, considering the coupling between quark degrees of freedom and gluon degrees of freedom. The PNJL model shows features of both chiral symmetry restoration and deconfinement phase transition, so this may allow the PNJL model to better describe the properties of QCD matter under rotation at high temperatures and finite chemical potentials. The PNJL model under rotation has been proposed in Ref. \cite{Sun:2023dwh}, thus, it is meaningful to calculate these thermodynamics in this model. Although, there is still controversy on how rotation affects the deconfinement transition at present, and we hope the lattice QCD provides more clues on the Polyakov loop, finally, make this research available in the PNJL model.

\section*{Acknowledgements}
We greatly thank Mei Huang, Kun Xu and Jie Mei for useful discussions. The work has been supported by the National Natural Science Foundation of China (NSFC) with Grant No. 12375137 and 12205309, the start-up funding from University of Chinese Academy of Sciences(UCAS), and the Fundamental Research Funds for the Central Universities. 

\appendix
\section{Thermodynamic quantities}
\label{app:thermodynamic}

We list the detailed expressions of the entropy density and quark number density, and the angular momentum along the $z$-axis:
\begin{widetext}
\begin{eqnarray}
s =&& \frac{3}{{4{\pi ^2}}}\sum\limits_f {\sum\limits_{n =  - \infty }^\infty  {\int_0^{\infty} {{{\rm{p}}_t}d{{\rm{p}}_t}\int_{ - \infty}^{\infty} {d{p_z}} } \left( {\left( {{J_{n + 1}}{{({{\rm{p}}_t}r)}^2} + {J_n}{{({{\rm{p}}_t}r)}^2}} \right)} \right.} } \\ \nonumber
&& \times T\left\{ {\frac{{{{\rm{e}}^{ - \frac{{{\varepsilon_{f,n}} - {\mu _f}}}{T}}}\left( {{\varepsilon_{f,n}} - {\mu _f}} \right)}}{{\left( {1 + {{\rm{e}}^{ - \frac{{{\varepsilon_{f,n}} - {\mu _f}}}{T}}}} \right){T^2}}} + \frac{{{{\rm{e}}^{ - \frac{{{\varepsilon_{f,n}} + {\mu _f}}}{T}}}\left( {{\varepsilon_{f,n}} + {\mu _f}} \right)}}{{\left( {1 + {{\rm{e}}^{ - \frac{{{\varepsilon_{f,n}} + {\mu _f}}}{T}}}} \right){T^2}}}} \right.\\ \nonumber
&&\left. { + \log \left[ {1 + {{\rm{e}}^{ - \frac{{{\varepsilon_{f,n}} - {\mu _f}}}{T}}}} \right] + \log \left[ {1 + {{\rm{e}}^{ - \frac{{{\varepsilon_{f,n}} + {\mu _f}}}{T}}}} \right]} \right\},
\end{eqnarray}

\begin{eqnarray}
n =&& \frac{3}{{4{\pi ^2}}}\sum\limits_f {\sum\limits_{n =  - \infty }^\infty  {\int_0^\infty  {{{\rm{p}}_t}d{{\rm{p}}_t}\int_{ - \infty}^{\infty} {d{p_z}} } \left( {\left( {{J_{n + 1}}{{({{\rm{p}}_t}r)}^2} + {J_n}{{({{\rm{p}}_t}r)}^2}} \right)} \right.} } \\ \nonumber
&&\times\frac{{{{\rm{e}}^{\frac{{2{M_f} + \omega  + 2n\omega }}{{2T}}}}\left( { - 1 + {{\rm{e}}^{\frac{{2{\mu _f}}}{T}}}} \right)}}{{\left( {{{\rm{e}}^{\frac{{{M_f} + {\mu _f}}}{T}}} + {{\rm{e}}^{\frac{{\left( {\frac{1}{2} + n} \right)\omega }}{T}}}} \right)\left( {{{\rm{e}}^{\frac{{{M_f}}}{T}}} + {{\rm{e}}^{\frac{{2{\mu _f} + \omega  + 2n\omega }}{{2T}}}}} \right)}},
\end{eqnarray}

\begin{eqnarray}
J =&& \frac{3}{{4{\pi ^2}}}\sum\limits_{n =  - \infty }^\infty  {\int_0^\Lambda  {{{\rm{p}}_t}d{{\rm{p}}_t}\int_{ - \sqrt {{\Lambda ^2} - p_t^2} }^{\sqrt {{\Lambda ^2} - p_t^2} } {d{p_z}} } \left( {\left( {{J_{n + 1}}{{({{\rm{p}}_t}r)}^2} + {J_n}{{({{\rm{p}}_t}r)}^2}} \right)} \right.} \left( { - 1 - 2n} \right)\\\nonumber
\\\nonumber
&& + \frac{3}{{4{\pi ^2}}}\sum\limits_f {\sum\limits_{n =  - \infty }^\infty  {\int_0^{\infty} {{{\rm{p}}_t}d{{\rm{p}}_t}\int_{ - \infty}^{\infty} {d{p_z}} } \left( {\left( {{J_{n + 1}}{{({{\rm{p}}_t}r)}^2} + {J_n}{{({{\rm{p}}_t}r)}^2}} \right)} \right.} } \\\nonumber
\\\nonumber
&& \times \frac{{\left( {{{\rm{e}}^{{\varepsilon_{f,n}}/T}} + 2{{\rm{e}}^{\frac{{{\mu _f}}}{T}}} + {{\rm{e}}^{\frac{{{\varepsilon_{f,n}} + 2{\mu _f}}}{T}}}} \right)\left( {1 + 2n} \right)}}{{2\left( {{{\rm{e}}^{{\varepsilon_{f,n}}/T}} + {{\rm{e}}^{\frac{{{\mu _f}}}{T}}}} \right)\left( {1 + {{\rm{e}}^{\frac{{{\varepsilon_{f,n}} + {\mu _f}}}{T}}}} \right)}}.
\end{eqnarray}
\end{widetext}

\bibliography{ref-lib}
\end{document}